\documentclass[sigconf]{acmart}
\usepackage{bm}
\usepackage{multicol}
\usepackage{caption}
\usepackage{subcaption}
\usepackage{algorithm2e}
\usepackage{algpseudocode}
\AtBeginDocument{%
  \providecommand\BibTeX{{%
    \normalfont B\kern-0.5em{\scshape i\kern-0.25em b}\kern-0.8em\TeX}}}

\newcommand{\highlight}[1]{\textcolor{black}{#1}}

\copyrightyear{2024}
\acmYear{2024}
\setcopyright{acmlicensed}
\acmConference[CHI '24]{Proceedings of the CHI Conference on Human Factors in Computing Systems}{May 11--16, 2024}{Honolulu, HI, USA}
\acmBooktitle{Proceedings of the CHI Conference on Human Factors in Computing Systems (CHI '24), May 11--16, 2024, Honolulu, HI, USA}
\acmDOI{10.1145/3613904.3642587}
\acmISBN{979-8-4007-0330-0/24/05}

\begin{document}
\SetKwComment{Comment}{/* }{ */}
\title[Surgment]{Surgment: Segmentation-enabled Semantic Search and Creation of Visual Question and Feedback to Support Video-Based Surgery Learning}
\author{Jingying Wang}
\email{wangchy@umich.edu}
\affiliation{%
  \institution{University of Michigan}
  \city{Ann Arbor}
  \state{Michigan}
  \country{USA}
}

\author{Haoran Tang}
\email{thr@umich.edu}
\affiliation{%
  \institution{University of Michigan}
  \city{Ann Arbor}
  \state{Michigan}
  \country{USA}
}

\author{Taylor Kantor}
\email{tkantor@med.umich.edu}
\affiliation{%
  \institution{University of Michigan}
  \city{Ann Arbor}
  \state{Michigan}
  \country{USA}
}

\author{Tandis Soltani}
\email{tandiss@med.umich.edu}
\affiliation{%
  \institution{University of Michigan}
  \city{Ann Arbor}
  \state{Michigan}
  \country{USA}
}

\author{Vitaliy Popov}
\email{vipopov@umich.edu}
\affiliation{%
  \institution{University of Michigan}
  \city{Ann Arbor}
  \state{Michigan}
  \country{USA}
}

\author{Xu Wang}
\email{xwanghci@umich.edu}
\affiliation{%
  \institution{University of Michigan}
  \city{Ann Arbor}
  \state{Michigan}
  \country{USA}
}

\renewcommand{\shortauthors}{Wang et al.}

\begin{abstract}
Videos are prominent learning materials to prepare surgical trainees before they enter the operating room (OR). In this work, we explore techniques to enrich the video-based surgery learning experience. We propose Surgment, a system that helps expert surgeons create exercises with feedback based on surgery recordings. Surgment is powered by a few-shot-learning-based pipeline (SegGPT+SAM) to segment surgery scenes, achieving an accuracy of 92\%. The segmentation pipeline enables functionalities to create visual questions and feedback desired by surgeons from a formative study. Surgment enables surgeons to 1) retrieve frames of interest through sketches, and 2) design exercises that target specific anatomical components and offer visual feedback. In an evaluation study with 11 surgeons, participants applauded the search-by-sketch approach for identifying frames of interest and found the resulting image-based questions and feedback to be of high educational value.

\end{abstract}


\settopmatter{printfolios=true}



\begin{teaserfigure}
   \includegraphics[width = \textwidth]{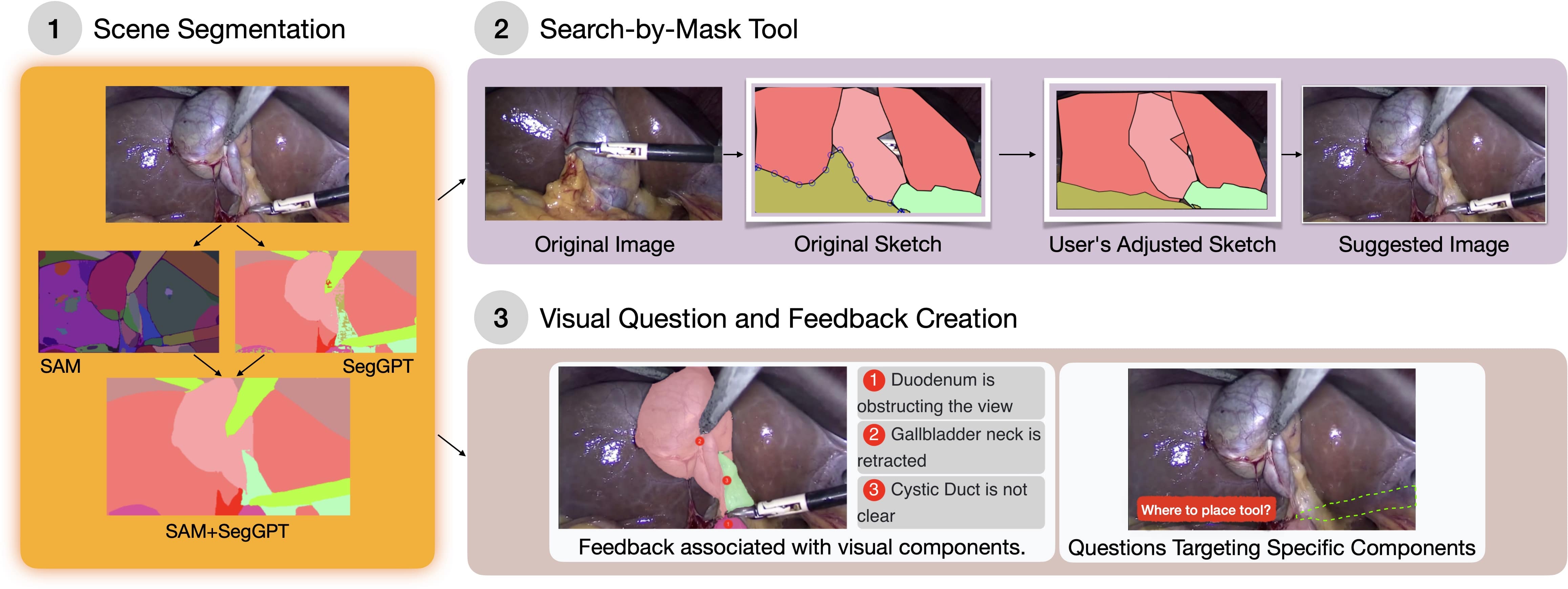}
    \caption{Overview of Surgment, a web-based system that helps expert surgeons create visual questions and feedback based on surgery videos to enhance video-based surgery learning. \textcircled{1} Surgment is powered by a surgery scene segmentation pipeline (SegGPT+SAM), which generates an accurate understanding of the surgery scene composition. Based on the scene segmentation result, Surgment has two key design features, namely \textcircled{2} A search-by-mask tool, which enables surgeons to quickly identify image frames by adjusting the position, size, and shape of the masks. \textcircled{3} A quiz-maker tool, which enables surgeons to create visual questions and feedback that target specific anatomical structures and surgical tools.}
  \label{fig:teaser}
  \Description[teaser figure, overview of Surgment]{On the left is the segmentation algorithm, on the top right is the search-by-mask tool, and on the bottom right is the extract a component and draw a path question type.}
\end{teaserfigure}

\maketitle

\section{Introduction}
Videos are prominent learning materials to prepare surgical trainees before they enter the operation room \cite{Singh2015, esposito2022video, mazer2022video}. However, the surgery video-watching experience has mostly been passive where the viewers have limited opportunities to actively engage with the video content and receive feedback \cite{esposito2022video, SNYDER2012643, keller2021video}. Prior work has explored techniques to enhance the interactivity of videos for learning, including supporting viewers to collaboratively create step-by-step descriptions \cite{weir2015learnersourcing} and concept maps \cite{liu2018conceptscape} for how-to videos, enabling teachers to embed multimedia exercises in lecture videos \cite{kim2015rimes}, and presenting the steps in surgery videos for viewers to search based on structures \cite{kim2023surch}. 

Learning of surgery videos, specifically laparoscopic surgeries, is a highly visual experience \cite{jva}. Surgical trainees need to acquire skills including analyzing the patient's anatomical structures, making operation decisions based on the patient's anatomy, and placing and using surgical tools. Prior work to enhance video learning experience is insufficient since they focus primarily on extracting the procedures in a task \cite{Fraser2020, Kim2014navigation, Nawhal2019, Pavel2014, Ponzanelli2019, Truong2021, Yang2022}, with little focus on helping viewers discern components within one image frame. Moreover, these techniques do not easily generalize to learning surgery videos, which involve subtle scene changes, such as the cleaning of a small vessel, in contrast with more dramatic scene changes in how-to videos \cite{Schoeffmann2015}. 

The literature on learning sciences theories \cite{chi2014icap}, online learning support systems \cite{shin2018understanding, xu2022elinor, koedinger2015learning}, and video-based surgery coaching practice \cite{esposito2022video, augestad2020video, daniel2022video, soucisse2017video} presented overwhelming evidence that embedding interactive questions and feedback during video watching improves learners' experience and learning outcomes. However, prior work on question generation are primarily text-based \cite{Wang2019, Yeckehzaare2020, Das2021, Kurdi2020, Wang2022, Wang2018, Majumder2014, Majumder2015, Shen2018OnTG, Yuan2019AutomaticGO, Zhu2021}. Automatic visual question generation techniques for educational purposes often produce simple questions, e.g., counting objects in a scene \cite{Xie2021, fan2018}. Techniques to generate visual questions that target higher-order thinking skills are limited. 

To enhance the learning experiences of trainees, we present Surgment to help create high-quality visual questions based on surgery videos. We started with a formative study aimed at understanding the challenges and needs of surgeons. We found that surgeons heavily rely on visual cues in videos to identify teachable moments, with the search for specific image frames being particularly time-consuming. They often annotated these images for both question creation and providing feedback. A key finding was the surgeons' interest in creating questions about anatomical structures, surgical tools, and operation procedures. Most importantly, the study highlighted that surgeons' unique knowledge and expertise in question and feedback creation are indispensable and should be carefully integrated into the system's design.

Building on these insights, we propose Surgment, a web-based system that helps expert surgeons create exercises with feedback based on surgery recordings. A core component of the Surgment system is an image segmentation pipeline (SegGPT+SAM), which aims to generate an accurate understanding of surgical scene compositions, making it possible to query the videos for specific surgery scenes and manipulate certain anatomical structures and tools. In developing the pipeline, we integrated two models, namely the Segment Anything Model (SAM) \cite{kirillov2023sam} which accurately differentiates regions, and SegGPT \cite{wang2023seggpt} which makes class prediction for the segmented regions. We found this approach (SegGPT+SAM) to be effective in compensating for the errors each model would make, as shown in \textcircled{1} Scene Segmentation of Fig.\ref{fig:teaser}.

Surgment has two unique design features to assist expert surgeons, inspired by the formative study findings.
First, the search-by-mask tool (Fig.\ref{fig:teaser} \textcircled{2}) enables surgeons to quickly identify image frames by expressing the desired requirement. Every recognized component on the scene is represented as an editable polygon. The user can adjust the position, size, and shape of the polygon to indicate a new composition of the frame they are interested in. For example, the user may adjust the polygon of ``fat'' to indicate that they want frames with ``less fat occluded to the gallbladder''. The search engine will fetch the images that meet the requirements. Second, the quiz-maker tool enables surgeons to peruse the images retrieved and create visual questions with feedback. Surgeons can easily highlight regions on the images to offer visual feedback associated with the anatomic components, as shown in Fig.\ref{fig:teaser} \textcircled{3}. Surgeons can also create questions that target specific anatomical structures or surgical tools, e.g., asking about the direction of the force that needs to be applied to a surgical tool (Fig.\ref{fig:teaser}\textcircled{3}).

We performed a two-part evaluation of Surgment: 1) a technical evaluation of the segmentation pipeline and the search-by-mask tool for retrieving images, and 2) an evaluation study with 11 expert surgeons. 

We first assessed the SegGPT+SAM pipeline on two public lap chole datasets \cite{hong2020cholecseg8k, maqbool2020m2caiseg}, where it achieved an F1-score of 0.92. This score surpasses the leading regression models, UNet \cite{ronneberger2015unet} and UNet++ \cite{zhou2018unet++}, which scored 0.73 and 0.76 respectively. Furthermore, we compared against the top few-shot learning model, SegGPT \cite{wang2023seggpt} alone, which had an F1-score of 0.84. This comparison highlights the efficacy of the integrated SegGPT+SAM approach over few-shot learning models like SegGPT alone. Secondly, our evaluation showed that the search-by-mask tool effectively retrieves images meeting user requirements, with an accuracy rate of 88\%, significantly outperforming the baseline \cite{Leibetseder2020} approach's 31.1\% accuracy.

We fully evaluated all the system components and probed into surgeons' attitudes towards the educational value of the questions created using Surgment in an evaluation study with 11 expert surgeons. 
The study validated surgeons’ need to enhance video-based learning. Surgeons appreciated that Surgment allowed them to express semantic requirements when searching for surgery scenes in teaching and considered the custom-made visual feedback on authentic surgery scenes to be better than diagrams in textbooks. All surgeons in the study were able to successfully create exercises and feedback that they considered to be of high educational value, which suggested that the user interactions in Surgment were effectively designed. Surgeons considered Surgment to be a nice complement to existing training approaches and could help trainees prepare before they enter the operating room. 
Surgeons also pointed out areas they would like to see improvement for future AI-assisted tools to support educational material creation. First, surgeons reported a learning curve when using Surgment and considered natural user interfaces using voice commands to be desirable. Second, surgeons reported that finer-grained segmentation could help them offer higher-quality feedback. Third, Surgment helps surgical trainees learn operation flow, anatomy, and tool handling, whereas communication, 3D perception, and in-situ decision-making skills need to be developed through hands-on experiences.

\section{Related Work}
We discuss related work on video-based learning, surgery teaching and learning, image-based question generation, and video navigation and image retrieval techniques. 

\subsection{Video-based Learning}
Prior work has explored techniques to enhance the interactivity of videos for learning. 
For example, Kim et al. visualized previous viewers' interaction data to help future viewers navigate \cite{kim2014data}.
ToolScape \cite{kim2014crowdsourcing, weir2015learnersourcing} augmented how-to videos with step-by-step structure and descriptions using a crowdsourcing workflow.
RIMES \cite{kim2015rimes} enables teachers to embed multimedia exercises in lecture videos, where students can record their responses using video, audio, and inking. Chronicle \cite{grossman2010} allows users to learn from the document revision videos. Users can playback the revision history of content of interest, accompanied by navigation of the revision events. 
ConceptScape \cite{liu2018conceptscape} generates and presents a concept map for lecture videos through a crowdsourcing pipeline. \cite{shin2018understanding} shows that prompting learners to answer reflective questions while watching videos could help them learn. Surch \cite{kim2023surch} visualizes the steps of surgery videos through graphs and enables viewers to search for videos based on the structure. Prior techniques primarily focus on extracting the procedures in a task from videos, with little focus on helping viewers discern components within one image frame, which is critical in surgery learning. 

\subsection{Surgery Teaching and Learning}
Recent work in surgical education emphasizes the use of asynchronous methods to observe trainees' performance and offer feedback, which can better prepare trainees before they enter the operation room \cite{DEDEILIA_2020, hu_2017, soucisse_2017, Ahmet_2018, Augestad2020, SELL2021}.
Video recordings are essential learning materials for the learning of anatomical landmarks and surgical techniques \cite{ABDELSATTAR2015, Mentis2020}. Schmitz et al. showed that video content was more effective than traditional textbooks \cite{Schmitz2021}. Matsuda et al. \cite{Matsuda2021} also pointed out that videos allow trainees to learn the case multiple times but require a long time to watch. 
However, there are still challenges that restrict students from using it frequently and efficiently. Snyder et al. \cite{SNYDER2012643} found in their survey that resident surgeons and medical students do not benefit from videos due to a lack of help in interpreting them. Esposito et al. \cite{esposito2022video} and Avellino et al. \cite{Avellino2021} reported that video-based coaching has a lot of potential in surgical education, but the time commitment is a significant barrier. Editing the videos to create effective learning materials can be time-consuming, and this can limit the use of video-based coaching in surgical education.

\subsection{Question Generation Techniques}

Decades of research in learning sciences have demonstrated that question-answering activities improve learning more than having learners passively watch videos\cite{Wang2021, Hicks2016, Joyner2016, Kulkarni2015, Sajjadi2016}, which sparked the research interest in automatic question generation in support of active learning. Most prior work focused on text-based question generation \cite{Wang2019, Yeckehzaare2020, Das2021, Kurdi2020, Wang2022, Wang2018, Majumder2014, Majumder2015, Shen2018OnTG, Yuan2019AutomaticGO, Zhu2021}, making them less optimal in surgery learning which places high emphasis on visual content. 
Studies on video-based question generation still rely on texts, including subtitles, captions, and transcripts of the videos \cite{Ma2019, Nittala2023}. Xie et al. \cite{Xie2021} and Fan et al. \cite{fan2018} use images as sources for generating questions, but the questions generated are limited by type and content. Most questions are factual questions, e.g., asking the number of people in a scene, with limited educational value. 

\subsection{Key Phase Identification to Support Video Navigation}
Numerous studies have explored approaches to chunk video and select keyframes in support of video navigation \cite{matejka2013, Fraser2020, Kim2014navigation, Nawhal2019, Pavel2014, Ponzanelli2019, Truong2021, Yang2022}. Some methods rely on text information in the voice-over to select keyframes. Others \cite{Nawhal2019, Ponzanelli2019} focus on capturing local peaks of average similarity across time stamps. Munzer et al. \cite{Muenzer2017} introduced EndoXplore, a video player that enables users to add keyframes to the panel or accept keyframes suggested by the system based on the similarity of ORB keypoint descriptors \cite{Schoeffmann2015}. 
Several studies have developed techniques to identify key phases to support video navigation \cite{Shinozuka2022, Kirtac2022, Primus2016, Golany2022, czempiel2021opera, Garrow2021, guedon2021deep, kitaguchi2020real}. Surch \cite{kim2023surch} uses procedural graphs to represent surgery videos and enables efficient search of videos from a bank. In Surgment, we build upon prior work and display the keyframes in a surgery recording to help surgeons easily navigate the videos.

\subsection{Medical Image Retrieval Based on Patterns}
Previous research primarily utilizes texture-level similarity for image retrieval \cite{Hegde2019, Cai2019, Lindvall2021}. Studies like Hegde et al. \cite{Hegde2019} and Cai et al. \cite{Cai2019} use neural features for pattern matching and image fetching, focusing on pathology images with rich texture features. Conversely, xPath \cite{gu2023improving} and NaviPath \cite{gu2023navipath} adopt quantitative criteria for identifying regions of interest in pathological images. In the surgical context, where images are more about anatomic components than textures, SurgXplore \cite{Leibetseder2020} and IMOTION \cite{rossetto2015imotion} allow searches based on color patterns. However, these methods don't account for the spatial relationships between components.

\section{Formative Study}
We conducted an IRB-approved formative study with four surgeons to understand their process of creating questions and feedback based on surgery videos. The surgeons' demographic information is shown in Tab.\ref{tab:formative}.
\highlight{The study aimed to understand} 1) Would attending surgeons use image-based questions to coach trainees? 2) What types of questions are surgeons creating and how do they identify question opportunities? 3) How do surgeons create feedback for the questions? 4) What are the challenges surgeons face during their question creation process?

\subsection{Procedure}

\begin{table}[h]
\begin{tabular}{ccccc}
\hline
ID  & Gender & Race      & Profession & \# Lap Chole Performed \\ \hline
FP1  & Male   & Asian     & Attending        & 800+                   \\
FP2  & Female & Caucasian         & Fellow           & 50+                    \\
FP3  & Male   & Caucasian & Fellow           & 50+                     \\
FP4  & Male   & Caucasian         & PGY-5            & 50+                     \\
\hline
\end{tabular}
\caption{Demographic information of participants in the formative study. }
\label{tab:formative}
\Description[Demographic information of the formative study.]{A table reports the Gender, Race, Profession Level, and number of Lap Chole performed for the four expert surgeons.}
\end{table}

This study focuses on laparoscopic cholecystectomy (lap chole) surgeries, ideal for teaching general surgery due to their standardized nature. Conducted via Zoom, participants engaged in two 1-hour sessions. They watched a 30-minute lap chole YouTube video \cite{v1, v2, v3, v4} and created questions that they thought would help trainees learn on Jamboard \cite{jamboard} without format restrictions. The interviews prompted reflection on the question creation process, including question identification, use of visual/textual cues, rationale, and distractor creation in multiple-choice questions. Discussions also covered challenges faced, target trainee populations, and skills assessed per the Operative Performance Rating System (OPRS) \cite{oprs}, like instrument handling, anatomy knowledge, problem-solving, and camera operation. Participants evaluated question difficulty and value. The interviews were transcribed and analyzed using affinity diagrams \cite{moggridge2007designing}.

\subsection{Findings}
\subsubsection{Videos are valuable learning resources for surgical trainees.} 
FP2 noted that based on the video, trainees can acquire basic techniques, anatomy knowledge, and practical tips, such as appropriate grasping locations and dissection timing. All participants agreed that quiz questions could better prepare novices for the operating room (OR). Since attending surgeons routinely pose questions to interns and residents during surgeries in the operation room (OR), the quiz questions offer novices a preview of the inquiries they may face in the OR. 

\subsubsection{Surgeons want to ask questions about critical portions of the surgery and rely on visual cues to identify them.}
Participants emphasized that every surgery type has distinct stages of heightened technical complexity. In the lap chole procedure, exposing the gallbladder and establishing the critical view of safety are the most critical stages. Mastery in these stages differentiates novices from experts and is imperative for patient safety. Participants noted that they relied on visual cues on the screen when they created the questions.
For example, the exposure of the gallbladder
is assessed by observing the amount of attached tissue, such as fat, duodenum, and omentum to the gallbladder.

\subsubsection{Surgeons create questions that target the identification of anatomy, procedural decision-making, and instrument handling skills.}
\textbf{Basic Anatomy of Lap Chole} All participants created questions that touched on fundamental lap chole anatomy. This included identifying structures like the gallbladder, cystic artery, and cystic duct. 
\textbf{Procedural Decision Making} All participants created questions targeting the evaluation of gallbladder exposure and the establishment of a critical view of safety. Such questions were designed to assess trainees' judgment of the optimal timing for dissection, thereby gauging their proficiency in decision-making in the OR.
\textbf{Instrument Handling} Another prevalent category of questions was about the selection, placement, and application of surgical tools. This involves ensuring that students opt for the right tool, position it accurately, and apply appropriate force. Notably, FP2 and FP4 employed a cross symbol to omit a specific tool from an image, indicating concealing the right answer from students.

\subsubsection{The questions surgeons create are highly visual and require students to contrast and discern components on the scene.}
\begin{figure}[h]
    \centering
    \includegraphics[width=0.3\textwidth]{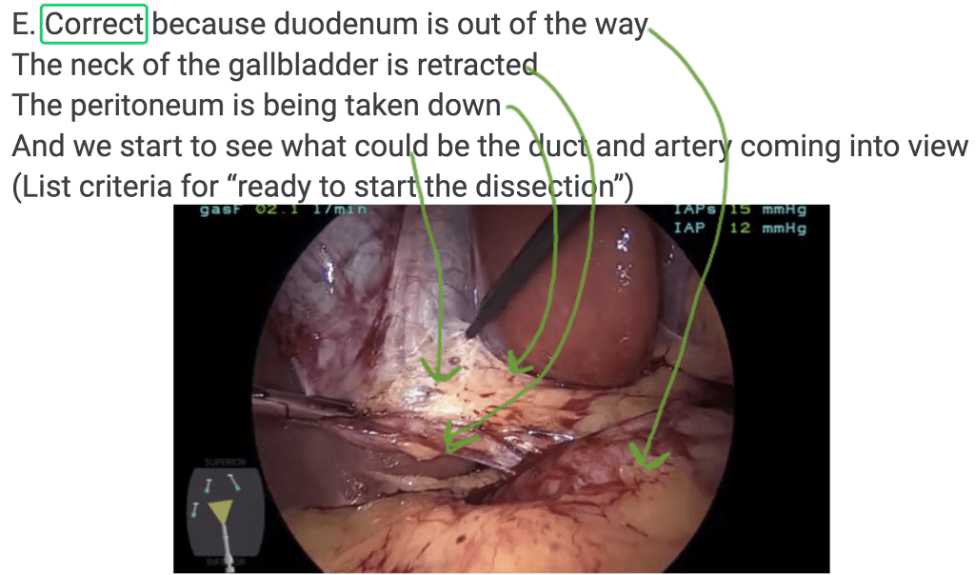}
    \caption{An example question and feedback provided by FP2 in the formative study. This is an option in a multiple-choice question they created asking ``When is enough exposure for dissection''. FP2 explained why this was the correct answer and linked the explanations with components of the surgery scene. }
    \label{fig:example}
    \Description[labeled feedback by surgeons]{A figure showing the surgical scene, with some labels linked to the gallbladder, cystic duct, etc.}

\end{figure}

All the questions created by surgeons are highly visual, involving careful differentiation of the anatomical components on the scene. 
Moreover, we noticed that surgeons often annotated images to give trainees feedback, e.g., as shown in Fig.\ref{fig:example}, FP2 used arrows to mark specific areas, indicating why this achieved sufficient exposure for dissection. In the questions that were designed to assess the motion and force of surgical tools, surgeons wanted trainees to illustrate the path for moving the surgical instrument or indicate the direction of the force to be applied to the surgical instrument. Additionally, it frequently occurred that FP1 and FP3 asked about the selection and placement of tools.

\subsubsection{Extracting images from videos requires the most effort and time from the experts.}
Experts dedicated time to meticulously reviewing the video, constantly moving forward and backward to select frames suitable for teaching. They based their choices on specific criteria. For instance, when FP2 created a multiple-choice question (as shown in Fig.\ref{fig:example}), they sought frames showing inadequate exposure for dissection, like those with excessive fat obscuring the gallbladder. Those questions aimed to help trainees discern various scenes, enhancing their decision-making skills in operations. When searching for frames, the surgeons had these specific criteria in their minds but they had to scroll through the video to identify such frames. This search process did not align with the surgeons' mental models.

\subsection{Design Goals}
Based on the formative study, we summarize the following design requirements for developing question-creation techniques to support video-based surgery learning. It is important to note that the formative study also revealed that surgeons possessed indispensable knowledge and expertise, which should be carefully incorporated into system design.

\begin{itemize}
    \item \textbf{D1. Support surgeons to quickly identify frames of interest.} The system should help surgeons quickly identify frames based on anatomical structures. Specifically, the system should help expert surgeons retrieve frames as contrasting cases for trainees to learn decision-making criteria in a procedure.
    \item \textbf{D2. Help surgeons navigate the videos to critical portions.}
    \item \textbf{D3. Enable surgeons to create questions that target specific visual elements on a scene.} 
    The system should help surgeons create questions that help trainees learn anatomy, surgical tool placement and motion, and decision-making skills. 
    \item \textbf{D4. Support surgeons to easily and conveniently offer feedback during question creation.} The system should help surgeons easily create feedback associated with visual elements on the scene.
\end{itemize}

\section{Surgment}

We developed Surgment, a web-based system for expert surgeons to create visual questions with feedback based on authentic surgery recordings. A core component of the Surgment system is an image segmentation pipeline (SegGPT+SAM), which aims to generate an accurate understanding of surgical scene compositions, making it possible to query the videos for specific surgery scenes and annotate the surgery scenes to highlight scene components like anatomical structures and tools. The segmentation pipeline enables the interaction designs in Surgment that help surgeons create exercises with feedback.

\begin{figure*}
    \centering
    \includegraphics[width = 0.95\textwidth]{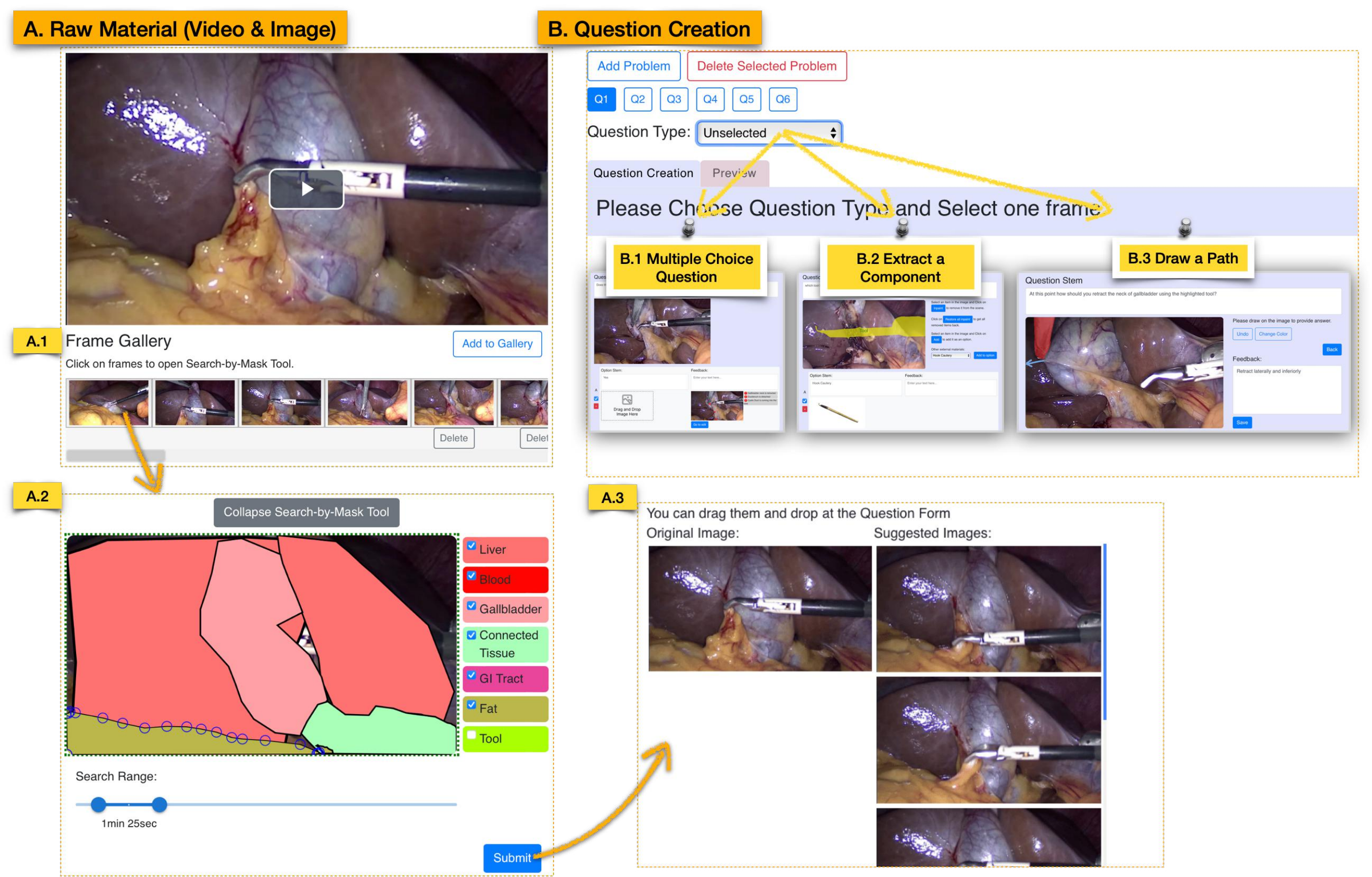}
    \caption{Overview of the Surgment platform. (A) The video panel displays an authentic lap chole surgery video, and the Frame Gallery (A.1) shows keyframes identified from the video. When a user clicks a keyframe, the search-by-mask canvas (A.2) expands for users to search for images by adjusting the size, shape, and positions of the polygon masks. The retrieved images are displayed to the right(A.3). (B) The Question Creation panel supports surgeons to create questions and feedback. Surgeons can select the images retrieved (A.3) in the previous step to create questions. Three question types are supported: MCQ (B.1), Extract a Component (B.2) and Draw a Path (B.2). }
    \label{fig:interface}
    \Description[web interface]{Web interface of Surgment.}
\end{figure*}

As shown in Fig.\ref{fig:interface}, the interface comprises two main sections: 1) A video panel with a keyframe gallery that displays key events from the video. When the user clicks an image from the keyframe gallery, the search-by-mask tool expands below. Every recognized component on the scene is represented as an editable polygon mask. The user can adjust the position, size, and shape of the polygon to indicate a new composition of the frame they are interested in. This mechanism enables surgeons to express their requirement for identifying an image (\textbf{D1, D2}). 2) A question creation panel that supports the creation of three types of image-based questions and feedback, namely ``Multiple Choice Question'' (Fig.\ref{fig:mcq}), ``Extract a Component'' (Fig.\ref{fig:mcqa}), and ``Draw a Path'' (Fig.\ref{fig:a}). Surgment enables surgeons to provide feedback associated with the components (e.g., ``duodenum'') in each scene (\textbf{D4}), and create questions targeting anatomical structures or surgical tools, e.g., how to place a surgical instrument (\textbf{D3}).

\subsection{SegGPT+SAM Surgery Scene Segmentation Pipeline}
The formative study points to the importance of generating an accurate understanding of surgery scenes for surgeons to teach based on the components. We designed and developed a scene segmentation pipeline combining the SegGPT \cite{wang2023seggpt} and the Segment Anything (SAM) models \cite{kirillov2023sam}. 

\subsubsection{Task Description}
The lap chole surgical scenes are comprised of various anatomical components and surgical tools. We conclude them into 9 categories: Background, Abdominal Wall, Liver, Gastrointestinal Tract, Fat, Grasper, Blood, Connected Tissue, and Gallbladder. The scene segmentation pipeline aims to label the regions in an image with one of the 9 classes. 

\subsubsection{Limitation of Regression Models}
Regression models are the most frequently used methods for scene segmentation in lap chole \cite{silva2022analysis}, with the UNet \cite{ronneberger2015unet} and UNet++ \cite{zhou2018unet++} models having the highest accuracy on surgery scene segmentation tasks. However, they have the following constraints:
\begin{enumerate}
    \item Limited generalizability and performance: Regression models require training on large public datasets which are often unavailable. Prior work showed that the UNet and UNet++ models trained on public dataset \cite{hong2020cholecseg8k} achieved a mean dice coefficient score (F1-score) of 0.54 and 0.62 \cite{silva2022analysis}.
    \item Insufficient differentiation capabilities: As illustrated in Fig.\ref{fig:connect-tool}, the segmentation results generated from UNet and UNet++ fail to distinguish between connected elements like the clip and the clip applicator. 
\end{enumerate}

\subsubsection{The Rationale of Combining SAM and SegGPT}
\begin{figure*}[h]
    \centering
    \includegraphics[width=\textwidth]{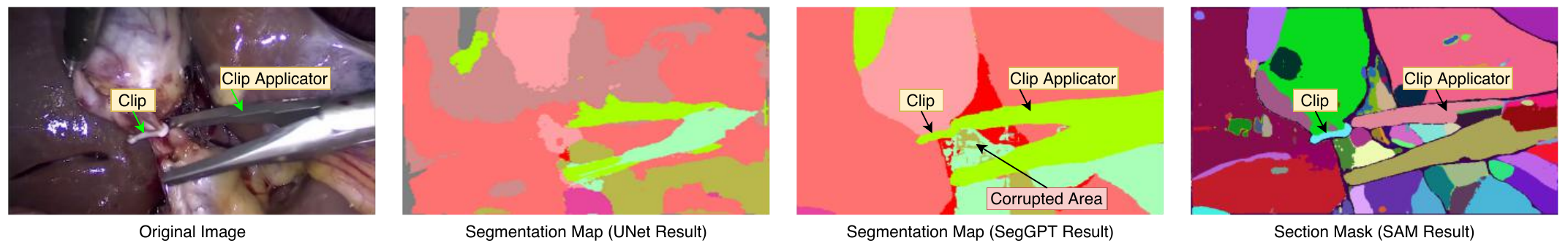}
    \caption{Segmentation result of UNet, SegGPT, and SAM on a single image. When two tools (clip and clip applicator) are adjacent in the image, the UNet and SegGPT models are not able to differentiate the two parts. SAM is able to distinguish the items, but cannot predict the classes of components.}
    \label{fig:connect-tool}
    \Description[segmentation result]{From left to right is the segmentation result of UNet, SegGPT, and SAM on a single image.}
\end{figure*}

We propose a scene segmentation pipeline combining SegGPT \cite{wang2023seggpt} and the ``Segment Anything'' model (SAM) \cite{kirillov2023sam}. We explain below why SAM or SegGPT alone is insufficient for surgical scene segmentation:
\begin{enumerate}
    \item SAM, although capable of dividing the image into sections, does not generate labels.
    \item The generated segmented output often includes numerous small sections, as shown in the ``SAM Result'' of Fig.\ref{fig:connect-tool}.
    \item SegGPT, similar to regression models, cannot distinguish between intersected, overlapped, or connected components in the same class, treating them as a unified entity instead of distinct parts, as demonstrated by the example of clip and clip applicator in Fig.\ref{fig:connect-tool}.
    \item SegGPT has prediction errors, as shown in the corrupted area in ``SegGPT Result'' of Fig.\ref{fig:connect-tool}.
\end{enumerate}

To adopt the strengths of both SAM and SegGPT, we propose a hybrid approach that leverages SAM's proficiency in image sectioning and SegGPT's capability in class prediction. This combined method aims to 1) merge some sections in the same class and 2) refine areas where the segmentation map corrupts.

\subsubsection{Propopsed Segmentation pipeline: SegGPT+SAM}
For a single frame $I$, the SegGPT model generates a map $M_{seggpt}$ for it, wherein the value at pixel $(x, y)$, belongs to one of nine categories $c_i$, mathematically represented as, 
\begin{equation}
    M_{seggpt}(x, y) \in \{c_i|i\in[0, 8]\}.
\end{equation}

\begin{figure*}[h]
    \centering
    \includegraphics[width=0.8\textwidth]{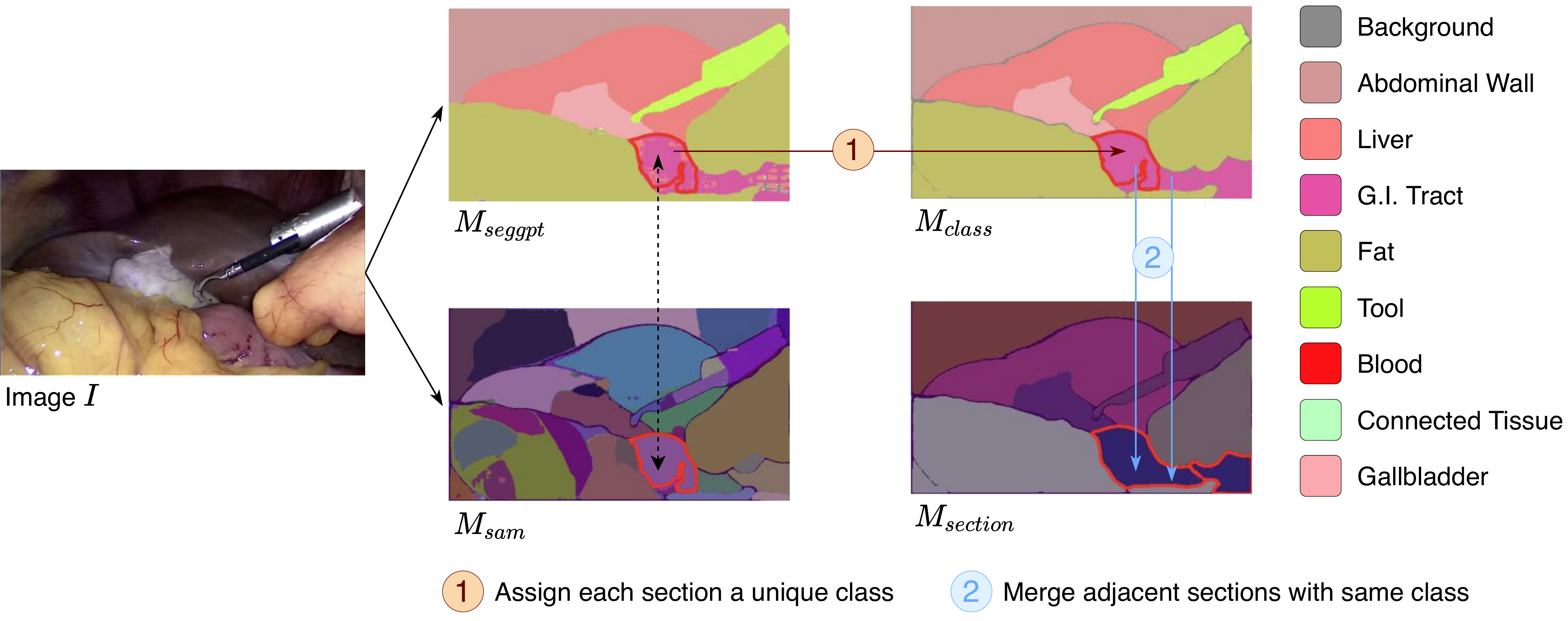}
    \caption{Our proposed SegGPT+SAM pipeline has two steps. 1) First, it assigns each section segmented by SAM to a unique class predicted by SegGPT. Since there is a discrepancy in the segmentation results achieved by the two models, we use a ``majority voting'' approach, selecting the class that the majority of the pixels in the section belong to. 2) Second, we merge the adjacent sections that have the same class.}
    \label{fig:sam+seggpt}
    \Description[A pipeline figure]{Two steps of our proposed SegGPT+SAM pipeline.}
\end{figure*}

Simultaneously, running the SAM model produces a mask $M_{sam}$, distinguishing $N$ individual sections within the scene. Consequently, the value at pixel $(x, y)$, pertains to one of these sections $s_j$, expressed as, 
\begin{equation}
    M_{sam}(x, y) \in \{s_j|j\in[0, N-1]\}.
\end{equation}
The pixel $(x,y)$ belongs either to a class $c_i$ or a section $s_j$. As depicted in Fig.\ref{fig:sam+seggpt}, our strategy is to merge some of the sections and assign each section a class, thereby constructing a mapping relation $\{s_i:c_j|i\in [0, N-1], j\in [0, 8] \}$. The algorithm consists of two steps. 
\begin{enumerate}
    \item{Step 1: assigning each section to a unique class.} For example, in Fig.\ref{fig:sam+seggpt}, the highlighted section (in red boundary) is recognized by SAM. However, from SegGPT, a small portion of the pixels in this highlighted section is mistakenly recognized as Liver, while the majority got the correct class, G.I. Tract. Therefore, we set all the pixels in this highlighted section to be G.I. Tract. When assigning a class to a section, we use a ``majority voting'' approach, selecting the class that the majority of the pixels in this section belong to. This method can effectively correct errors made by the SegGPT model.
    \item{Step 2: merge the sections that have the same class.} In Fig.\ref{fig:sam+seggpt}, both of the two sections highlighted have the class G.I. Tract. We will merge them. 
\end{enumerate}

With this method, we get a segmentation map $M_{class}$ that has class labels for every pixel, and section masks $M_{section}$ that differentiate the components on the scene. 

\subsubsection{Data labeling in SegGPT}
Since SegGPT is a few-shot learning method, it requires data labeling. In our implementation, we found that for a 10-minute video sampled at 25 fps, yielding 15,000 image frames, labeling 22 image frames (which is ~0.15\% of the total number of frames) gives us reasonable segmentation output.

\subsection{Video Panel}
The Video Panel in Surgment displays a surgery recording of the user's choice and the keyframes in the video. The keyframes are computationally identified based on visual features using a standard keyframe identification algorithm. Users can add new frames to the gallery. For a selected frame, users can use the search-by-mask tool (A.2) to retrieve similar frames that meet their criteria, as shown in Fig.\ref{fig:interface}.

\subsubsection{Key Frame Identification (A.1)} 
\begin{figure}[h]
     \centering
     \begin{subfigure}[b]{0.12\textwidth}
         \centering
         \includegraphics[width=\textwidth]{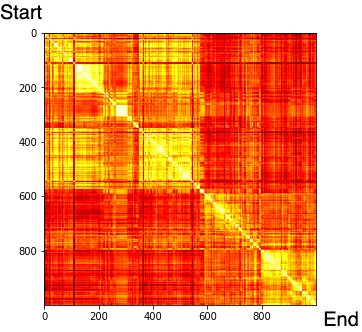}
         \caption{The similarity matrix.}
         \label{fig:similarity}
     \end{subfigure}
     \begin{subfigure}[b]{0.35\textwidth}
         \centering
         \includegraphics[width=\textwidth]{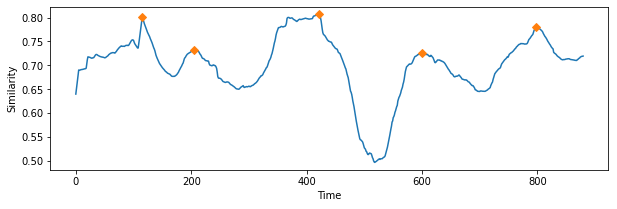}
         \caption{The local average similarity score.}
         \label{fig:signal}
     \end{subfigure}
        \caption{The light patches in the similarity matrix represent segments in the video that are visually consistent (a). We select the center for each patch as the keyframes, which are the peaks in the local average similarity score diagram (b). }
        \label{fig:keyframe}
        \Description[keyframe extraction]{The Left image is the similarity matrix with light patches along the diagonal axis and the right one is a signal along the time axis, going ups and downs.}
\end{figure}

We extracted keyframes in the video to support easy navigation following methods applied in prior literature \cite{keyframe_2005, Nawhal2019}. This process involves extracting features from each frame using pre-trained deep neural networks like VGG19's FC-4096 layer \cite{simonyan2014very}. Next, we calculated the cosine similarity between the features of all frame pairs. Fig.\ref{fig:similarity} visualizes the similarity matrix. The light patches represent visually consistent segments. We selected the center for each patch as the keyframes. We applied a mean filter along the diagonal axis of the matrix, which reflects the local average similarity as shown in Fig.\ref{fig:signal}. The peaks of this signal correspond to the patch centers.

\subsection{Search-by-Mask Tool}
Corresponding to Design Goal D1, we designed the search-by-mask tool, allowing users to identify surgery scenes by modifying polygon masks that represent components on a scene, as shown in Fig.6. Users select an image from the image frame gallery (A1), which opens the Search-by-Mask canvas with polygons generated based on the segmentation outputs (A2). The user can make adjustments to each polygon, including resizing, moving, rotating, and adjusting the shape by dragging the vertices of the polygon. The adjusted polygon masks are used to search for frames that match the scene composition. For example, if the user wants to identify frames with ``less fat occluded to the gallbladder'', they can adjust the polygon masks accordingly, as shown in \textcircled{2} the search-by-mask tool of Fig.\ref{fig:teaser} and Fig.\ref{fig:interface}. Based on the user-adjusted polygon masks, the system will run the search algorithm and provide nine suggested frames for the user to choose from (A.3).

\subsubsection{Generating polygons based on the segmentation maps}
The segmentation maps produced from the SegGPT+SAM pipeline enable us to differentiate the components on the scene. 
For classes ``G.I. Tract'', ``Fat'', ``Blood'', and ``Tools'', they may have multiple pieces that should appear separately on the scene. We sample the contours of the sections and render them separately. For classes ``Liver'' and ``Gallbladder'', there may also be multiple sections since an object such as a surgical tool may fragment the piece. But the sections should appear in one whole piece in the polygon masks. In this case, we will extract the alpha shape of all of the sections combined, to render them as one single piece. The contours of the components are used to generate polygons which are layered onto the search-by-mask canvas, following the order of ``Liver'', ``Gallbladder'', ``Fat'', ``G.I. Tract'', ``Blood'', and ``Tools'', as shown in Fig.\ref{fig:interface}.

\subsubsection{Search algorithm}
We identify frames whose segmentation maps are closest to the user's specification, as determined by the Mean Squared Error (MSE). Once the users finish adjusting the shapes and layout of the polygons, a reference map is generated based on the adjusted polygon masks to search for image frames in the video that have similar segmentation maps to the reference map. 
\begin{figure}[h]
     \centering
     \begin{subfigure}{0.5\textwidth}
         \centering
        \includegraphics[width = \textwidth]{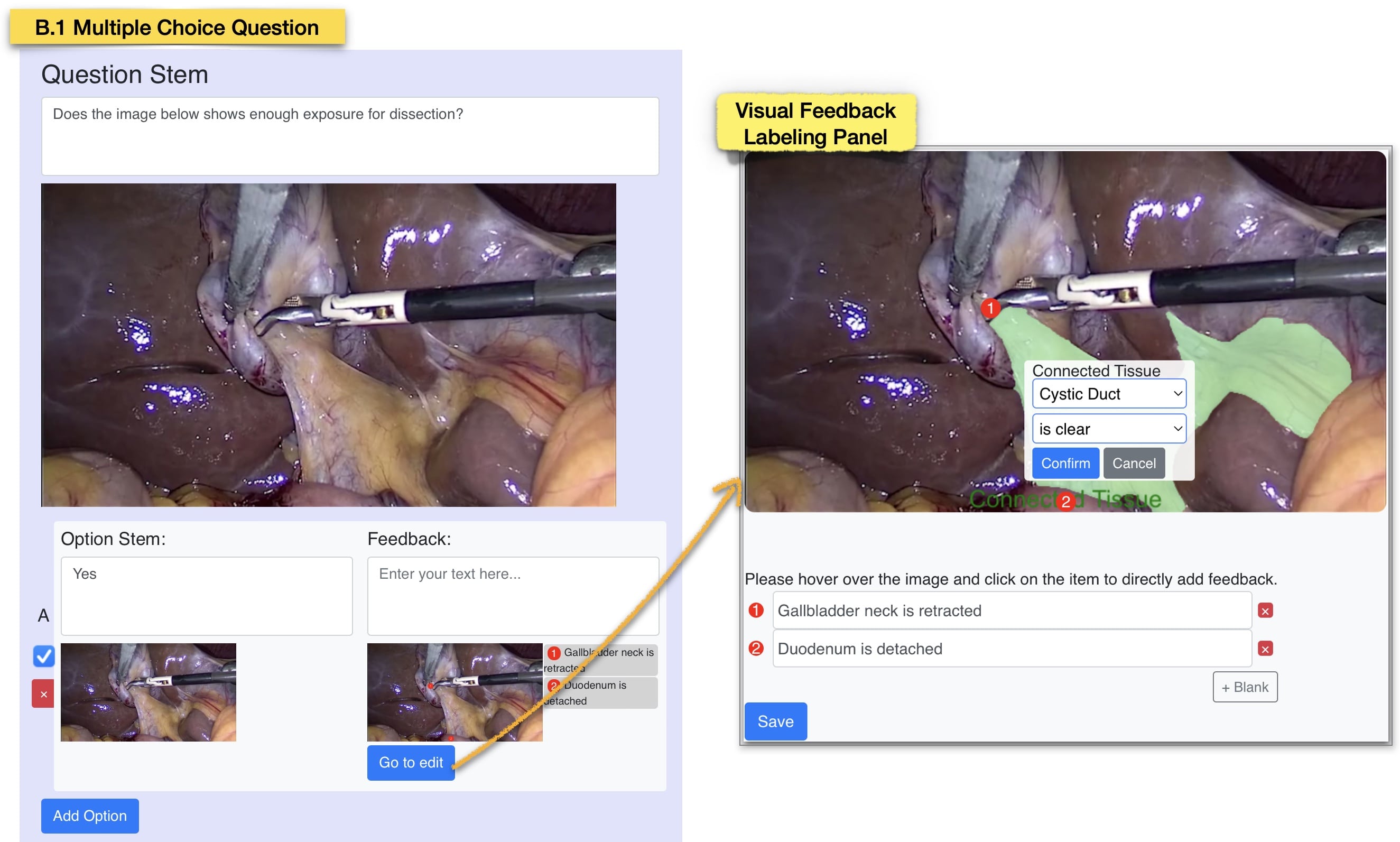}
        \caption{Users can insert text and images in both question stems and options. Users can highlight regions on the image to offer feedback. Surgment gives sentence starters for feedback, while surgeons can start from scratch as well.}
         \label{fig:mcq-s}
     \end{subfigure}
     \vfill
     \begin{subfigure}{0.5\textwidth}
         \centering
         \includegraphics[width=0.7\textwidth]{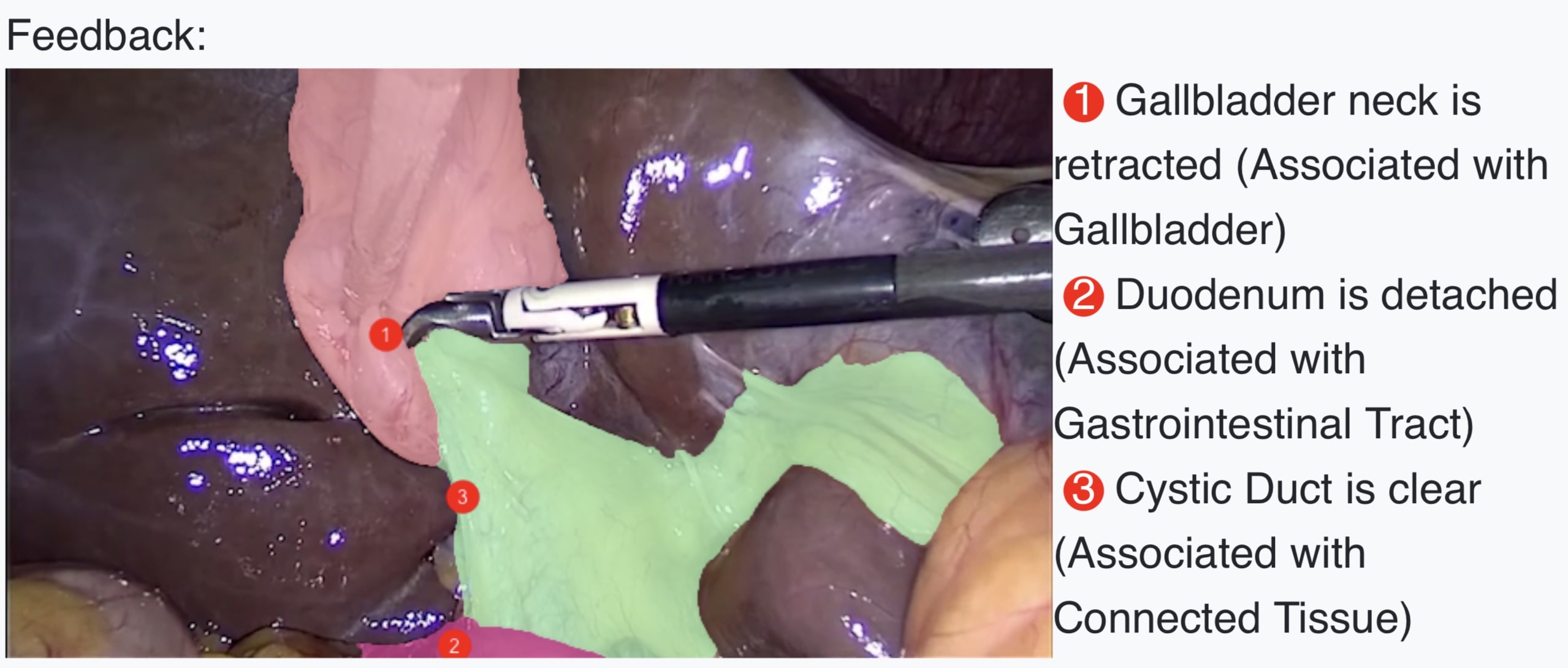}
         \caption{Preview of the feedback that students will receive}
         \label{fig:mcq-p}
     \end{subfigure}
        \caption{Multiple Choice Question.}
        \label{fig:mcq}
        \Description[Multiple Choice Question.]{A window on the web interface that highlights parts of the images, and a table of two figures accompanied with annotation is shown to the students. }
\end{figure}

\subsection{Question Creation Panel}
Users can create questions using the images in the frame gallery and the images retrieved from the search-by-mask tool. We implemented three question types based on surgeons' needs as evidenced in the formative study (D3, D4).

\textbf{Multiple-choice questions (B.1)} are the most common type of questions the surgeons created in the formative study. Surgeons expressed the importance of trainees comparing different scenes in a procedure, such as deciding when to start dissection based on gallbladder exposure. In Surgment, we enable the creation of image-based multiple-choice questions, with images in both the question stem and options. Aligning with Design Goal D4, surgeons can add feedback linked to specific scene components. As users hover over an image, segmented components are highlighted, allowing for feedback attachment to specific regions, as illustrated in Fig~\ref{fig:mcq}. Surgment thus facilitates high-quality visual feedback for students, as previewed in Fig.\ref{fig:mcq-p}.

\begin{figure}[h]
     \centering
     \begin{subfigure}{0.5\textwidth}
         \centering
        \includegraphics[width=0.7\textwidth]{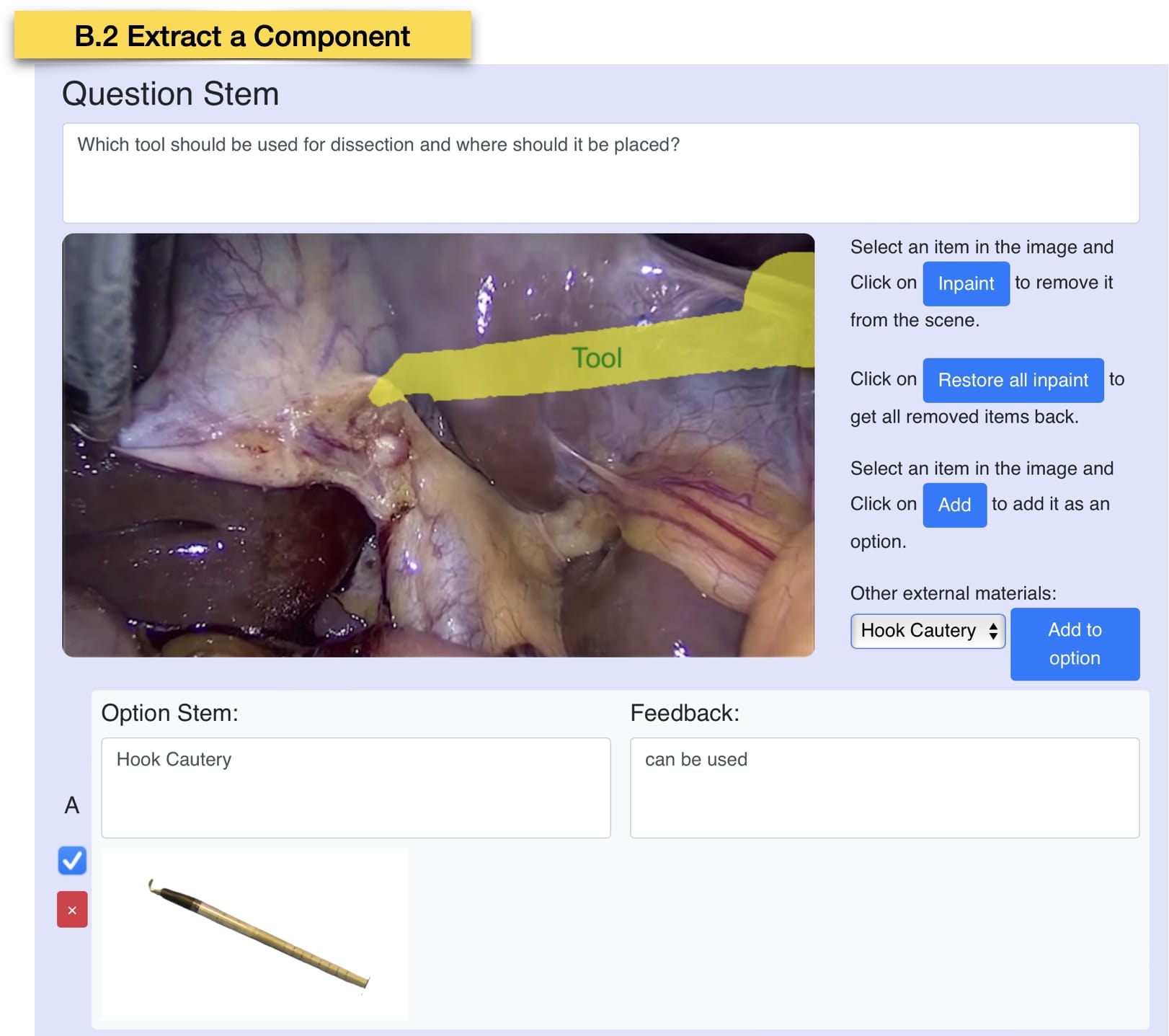}
        \caption{Users can select an item to remove it through ``Inpaint'' and get the removed items back through ``Restore all inpaint''. Users can select images from a bank to use as question options.}
         \label{fig:mcqa-s}
     \end{subfigure}
     \vfill
     \begin{subfigure}{0.5\textwidth}
         \centering
         \includegraphics[width=0.6\textwidth]{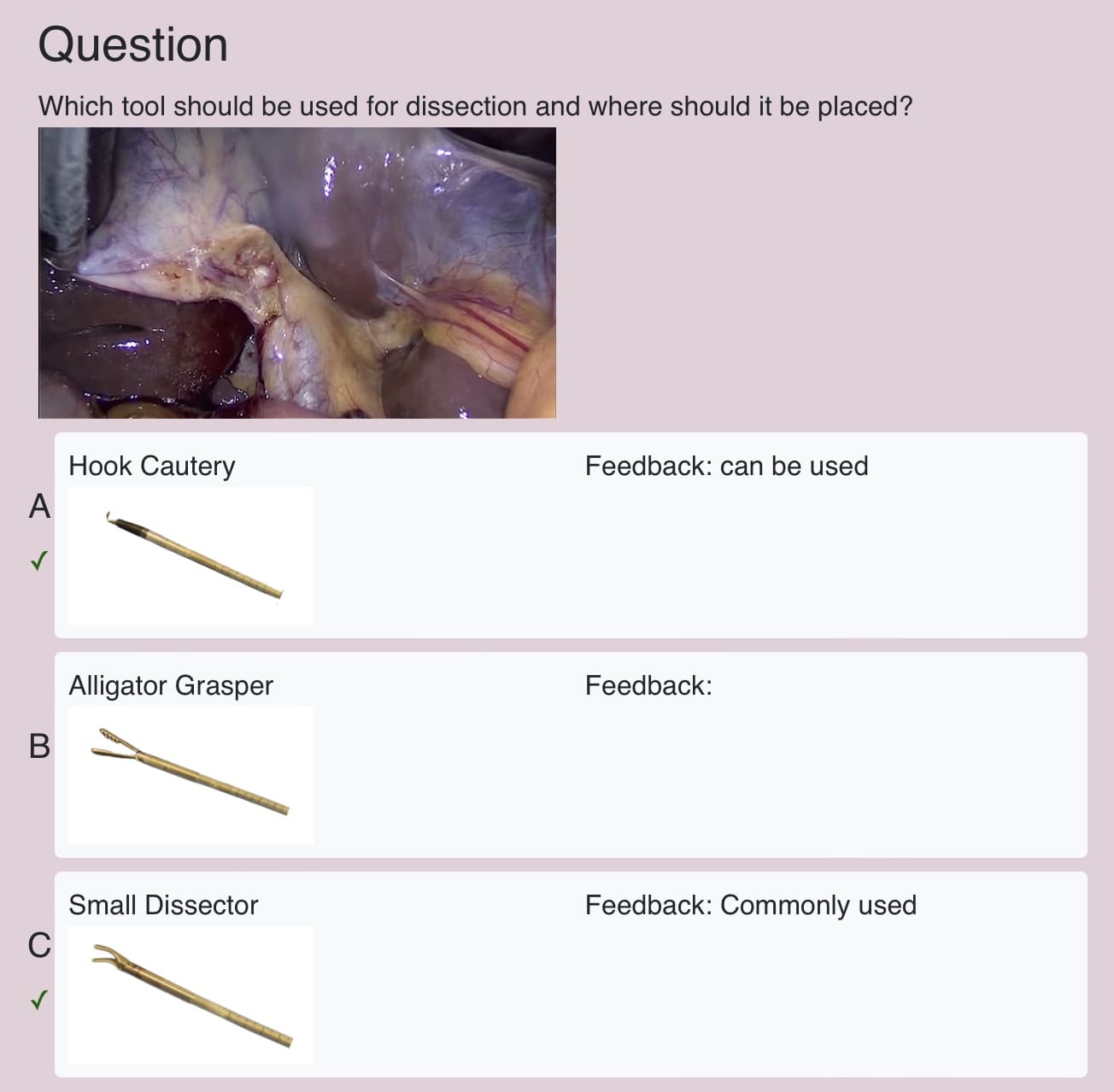}
         \caption{Preview of an ``Extract a Component'' question as a student. }
         \label{fig:mcqa-p}
     \end{subfigure}
        \caption{``Extract a Component'' Question.}
        \label{fig:mcqa}
    \Description[Extract a Component.]{A window showing a tool is removed from the image by AI model, and an interface that the student will see that the tool is missing. }
\end{figure}

Aligned with Design Goal D3, we introduced two question types in Surgment. The first, ``\textbf{Extract a Component (B.3)},'' allows users to select and remove a component from the scene using the ``Inpaint'' feature, accessible through the cloud API \cite{cleanuppictures}. This type enables surgeons to pose questions like ``Which tool should be utilized for dissection, and where should it be placed?'' The second type, ``\textbf{Draw a Path (B.3)},'' lets users choose a target component and invite respondents to draw a path, facilitating questions such as ``Which direction should you retract the neck of the gallbladder using the highlighted tool?''.

\begin{figure}[h]
     \centering
     \begin{subfigure}{0.5\textwidth}
         \centering
        \includegraphics[width = 0.7\textwidth]{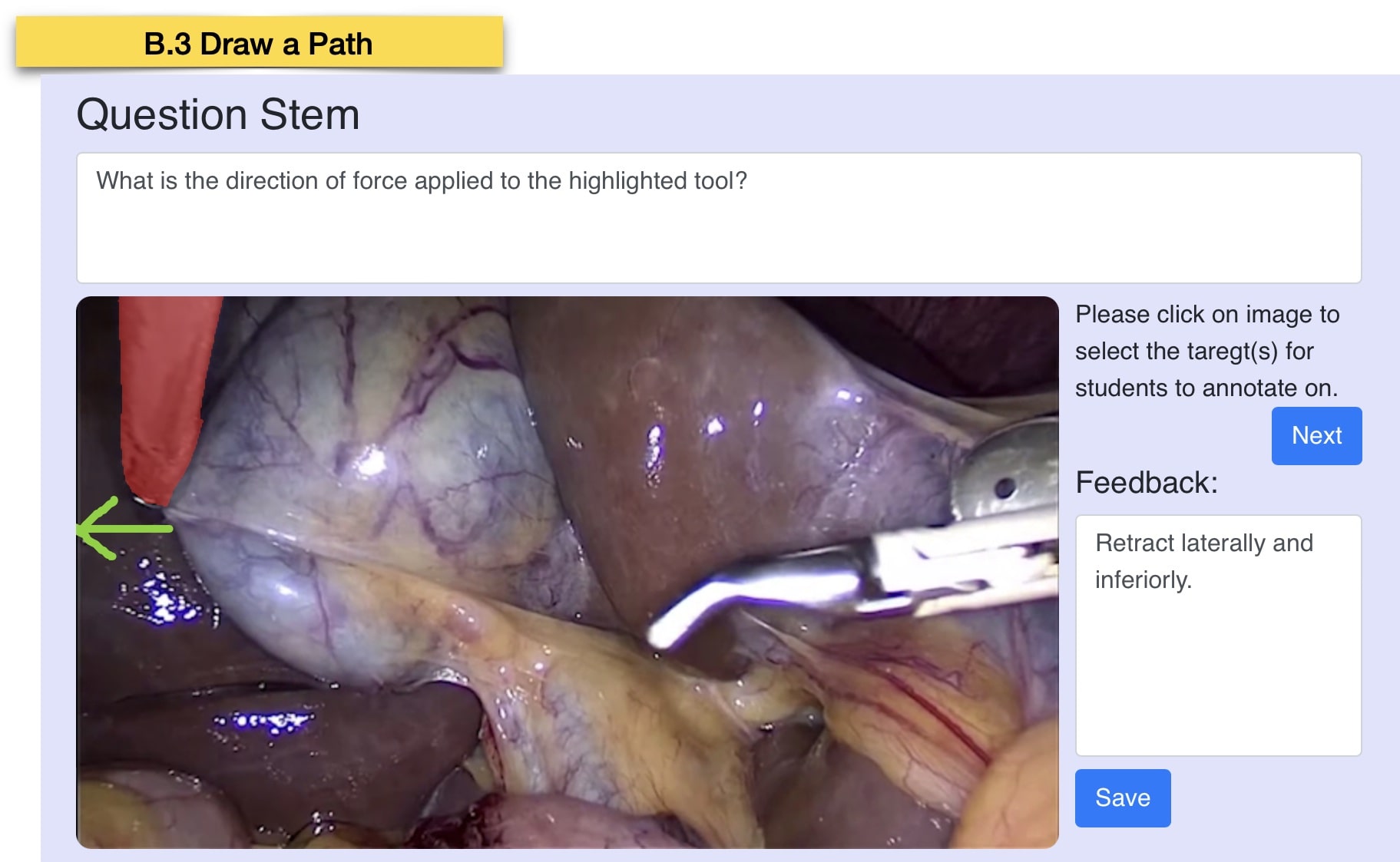}
        \caption{Users need to first select a target component they want to ask this question about, e.g., the surgical tool on the top left, and then draw a path as the correct answer.}
         \label{fig:a-s}
     \end{subfigure}
     \vfill
     \begin{subfigure}{0.5\textwidth}
         \centering
         \includegraphics[width=0.4\textwidth]{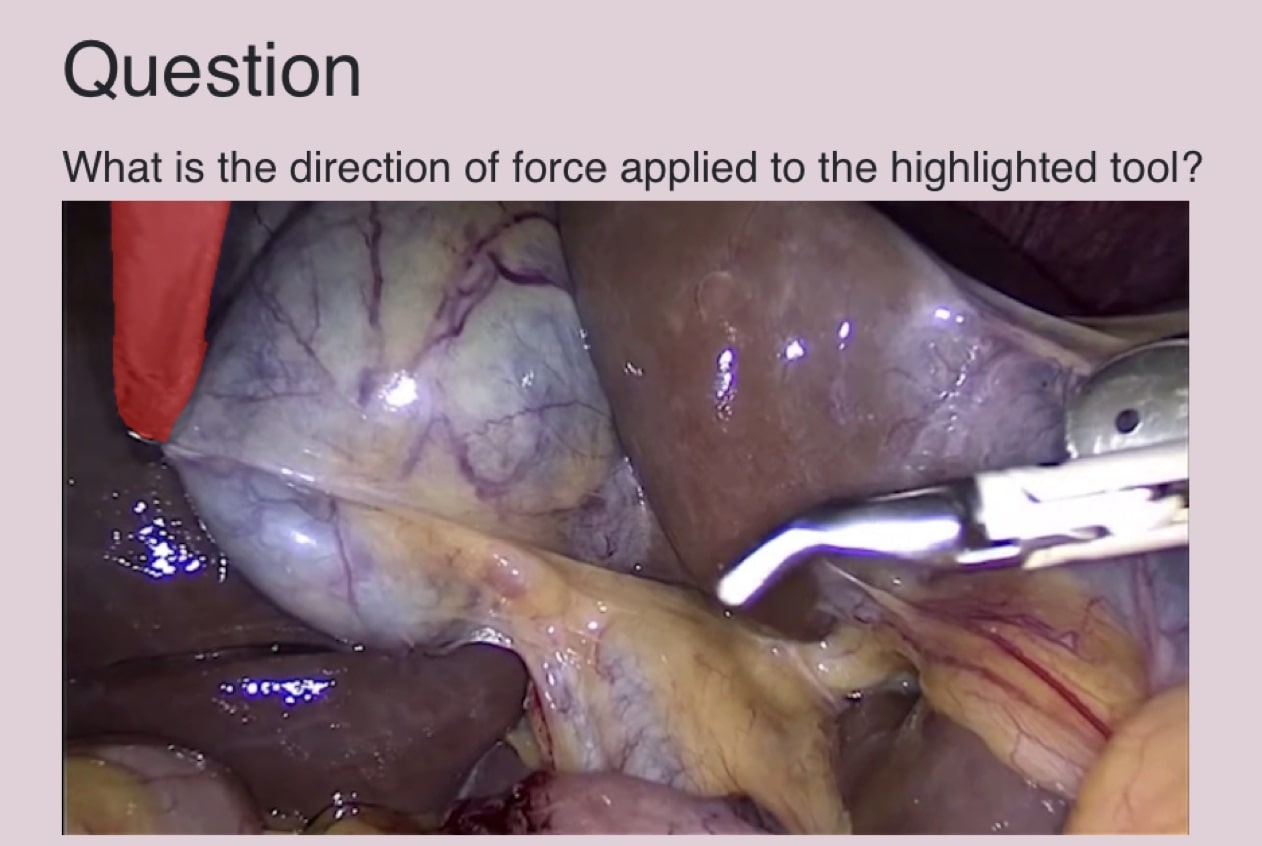}
         \caption{Preview of a ``Draw a Path'' as a student.}
         \label{fig:a-p}
     \end{subfigure}
        \caption{``Draw a Path'' Question.}
        \label{fig:a}
     \Description[Draw a path.]{A window shows that the user will be able to make a tool red and have a brush to draw on the image, and the student will see the red tool as the question, and the drawn path as the answer. }
\end{figure}

\subsection{Implementation}
In developing the front-end interface, we used several Javascript libraries, including react.js \cite{reactjs}, 
 fabric.js \cite{fabricjs}, firebase \cite{firebase}, etc.
The back-end server is built with the Python Django framework, which stores the surgery scene segmentation results. The system design and development went through several iterations. We have incorporated feedback from usability tests to improve the system design.

\section{Technical Evaluation}

\subsection{Technical Evaluation of the SegGPT+SAM Segmentation Pipeline}
We used the dice coefficient score (F1-score) as the metric for evaluating the segmentation result. A score of 0 denotes completely dissimilar masks, while a score of 1 denotes identical masks.
In current literature \cite{silva2022analysis}, regression models are the mainstream method for surgery scene segmentation. We chose UNet and UNet++, the most widely used regression models as the baselines. We trained UNet \cite{ronneberger2015unet} and UNet++ \cite{zhou2018unet++} on a merged dataset comprised of CholecSeg8k \cite{hong2020cholecseg8k} and m2caiSeg \cite{maqbool2020m2caiseg}. CholecSeg8k and m2caiSeg are the only public scene segmentation datasets for lap chole surgery images, with 13 and 19 classes respectively. Given that both datasets are limited in size, we created a combined dataset by synchronizing the labels between them. This involved merging classes into unified categories. For instance, we merged all the classes of instruments including ``Grasper'', ``Scissors'', ``Hook'', and ``Bipolar'' in the m2caiSeg dataset into the ``Tool'' class. 
Moreover, we selected a third baseline SegGPT \cite{wang2023seggpt}, which is a top-performing few-shot learning model for scene segmentation. The comparison of the SegGPT+SAM pipeline with SegGPT alone will help evaluate to what extent the combined approach improves over state-of-the-art few-shot learning methods. 
We evaluated all models on the combined dataset, which has not been done in the literature.

\begin{figure}[h]
    \centering
    \includegraphics[width=0.45\textwidth]{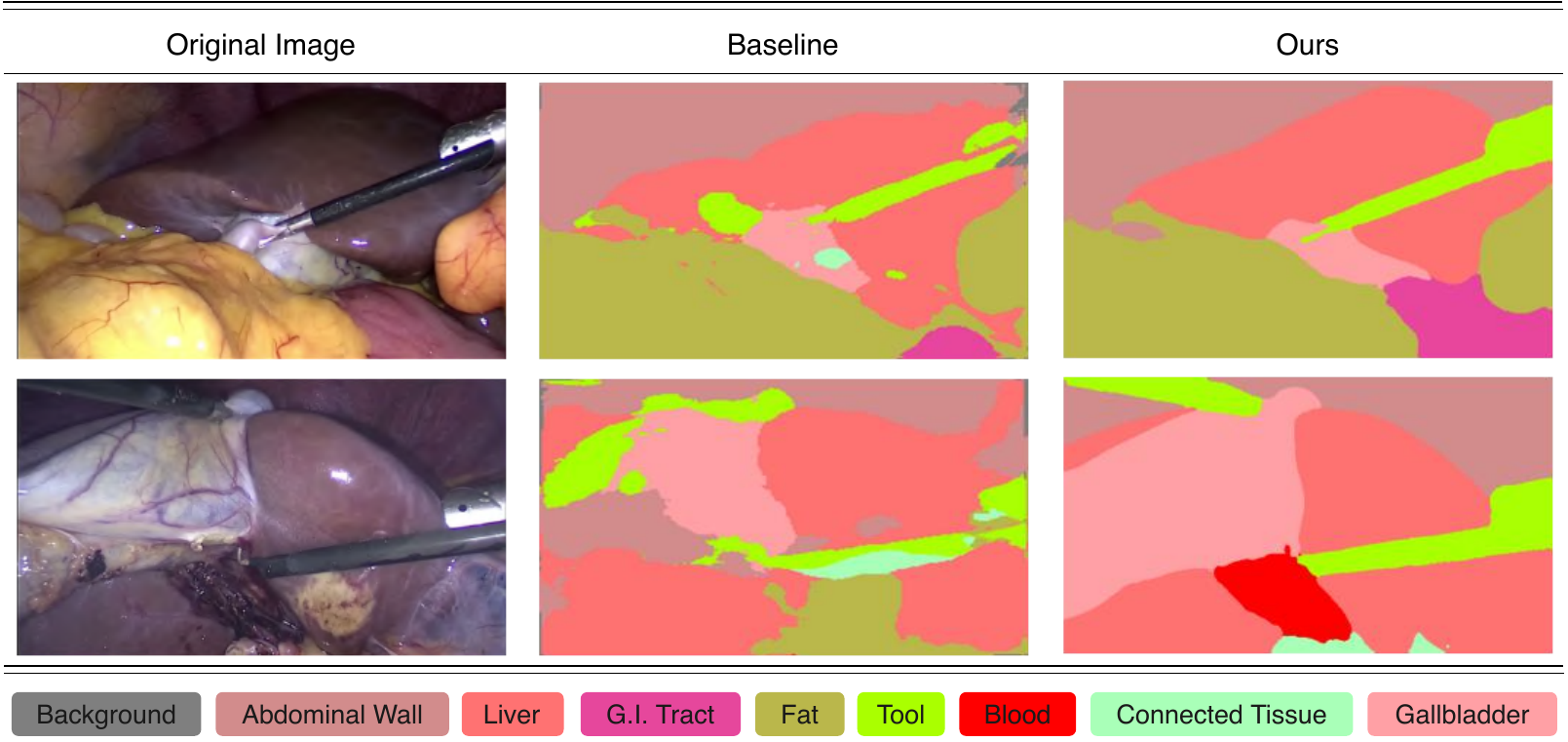}
    \caption{Example segmentation outputs generated by UNet (baseline) and SegGPT+SAM (Ours).}
    \label{fig:seg-compare}
    \Description[segmentation result.]{UNet can hardly recognize areas of gallbladder and tools, but our method can do it quite well. }
\end{figure}

The training set contains a total of 21427 labeled surgery scenes from surgery recordings. 
We took 22 labeled images, which is ~0.1\% of the training data as the source for SegGPT. 
The test set contains 782 images. We acknowledge that the size of the test set is a limitation. The scarcity of large datasets is a known challenge in medical tasks. As shown in Tab.\ref{tab:tdsc}, the baseline method UNet achieves an average dice coefficient score of 0.73, UNet++ achieves 0.76, and SegGPT achieves 0.84. 
Our SegGPT+SAM approach achieves 0.92, outperforming all three baselines, with a 21\% improvement over the regression models, and 9\% over SegGPT alone. Example segmentation outputs generated by the baseline UNet and our method are shown in Fig.\ref{fig:seg-compare}.

\begin{table*}[h]
\begin{tabular}{lcccc}
\hline
Class                  & Baseline (UNet)  & Baseline (UNet++)   & Baseline (SegGPT alone)     & Ours (SegGPT+SAM) \\ \hline
Background             & 0.90  & 0.98     & 0.68       & 0.98              \\
Abdominal Wall         & 0.52  & 0.32     & 0.86       & 0.84              \\
Liver                  & 0.69  & 0.69     & 0.86       & 0.90              \\
Gastrointestinal Tract & 0.53  & 0.68     & 0.85       & 0.95              \\
Fat                    & 0.57  & 0.76     & 0.86       & 0.84              \\
Tool                   & 0.70  & 0.77     & 0.85       & 0.91              \\
Blood                  & 0.99  & 0.99     & 0.86       & 0.98              \\
Connected Tissue       & 0.97  & 0.97     & 0.86       & 0.99              \\
Gallbladder            & 0.70  & 0.66     & 0.86       & 0.93              \\
Mean                   & 0.73  & 0.76     & 0.84       & 0.92              \\ \hline
\end{tabular}
\caption{Dice coefficient score (F1-score) of UNet, UNet++, SegGPT, and SegGPT+SAM.}
\label{tab:tdsc}
\Description[F1-score]{The F1 score of UNet is 0.73, UNet++ 0.76, SegGPT 0.84, and our method, SAM+SegGPT 0.92.}
\end{table*}

\begin{figure*}[h]
    \centering
    \includegraphics[width=\textwidth]{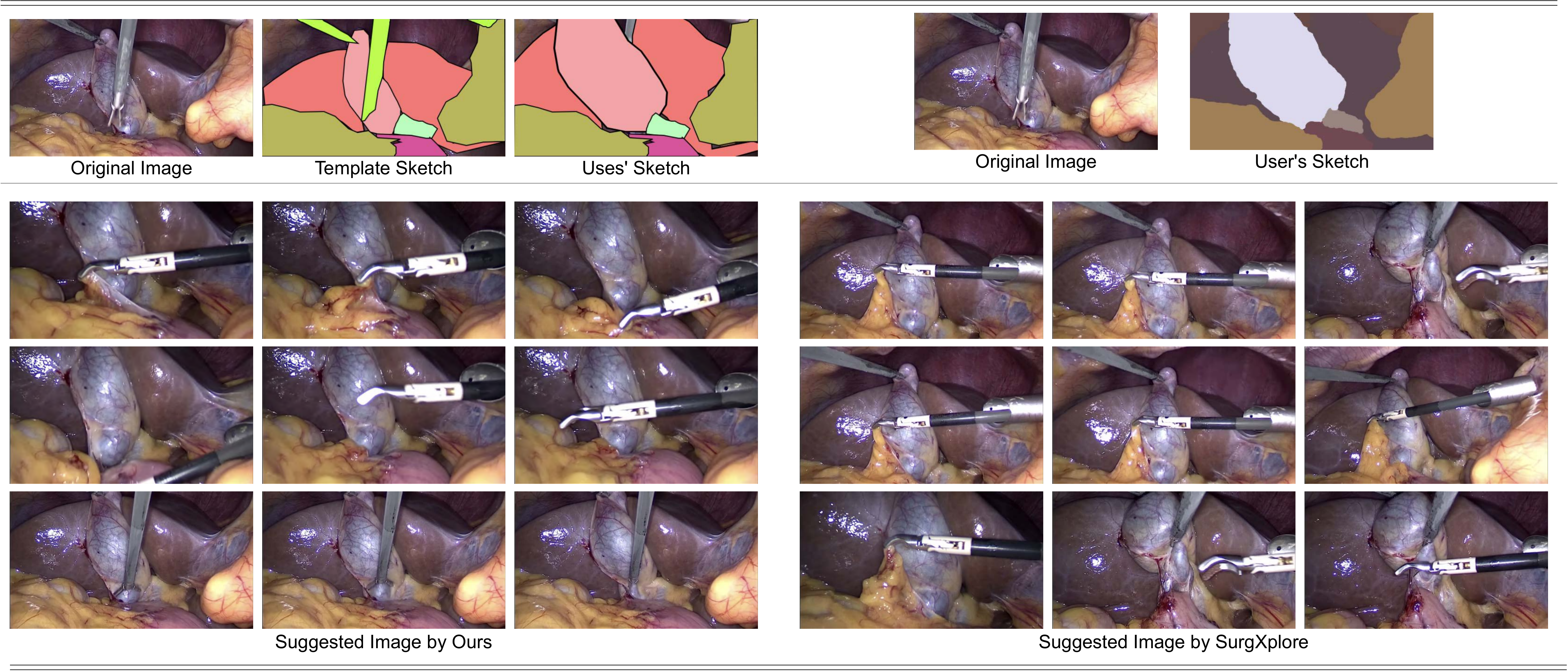}
    \caption{The sketch-based searching result of SurgXplore (baseline) and the Search-by-Mask (ours) based on the syntax of ``zoomed-in camera view, and focus on the bottom of the gallbladder''. As the view zoomed in, the gallbladder will look larger.}
    \label{fig:r1}
    \Description[SurgXplore v.s. Search-by-Mask]{A table showing results of our method on the left and SurgXplore on the right. Each method shows 9 suggested frames. }

\end{figure*}

\subsection{Technical Evaluation of the Search-by-Mask Tool}
The search-by-mask tool contributes a novel interaction method that provides segmentation maps for users to express their image retrieval requirements. Different from existing methods on image retrieval \cite{Dey_2019_CVPR, Yelamarthi_2018_ECCV, Bhunia_2020_CVPR}, Surgment's search-by-mask tool enables users to adjust the shapes and positions of the components in an image scene. 
The technical evaluation aims to evaluate the search-by-mask tool as an interaction design technique. It is worth noting that the search algorithm is not a contribution of the paper.
We evaluated the effectiveness of the search-by-mask tool in retrieving relevant images with a baseline method SurgXplore \cite{Leibetseder2020}.  
SurgXplore allows users to create free drawings as the basis used for retrieving images that match the pattern. It is the only study in the literature that retrieves endoscopic images utilizing user-generated templates. SurgXplore uses HistMap \cite{Loko2019} descriptor to measure the similarity between the user's sketch with the image frames in a video.


We adopted the evaluation metric frequently used in prior literature \cite{li2018survey, Dey_2019_CVPR, Yelamarthi_2018_ECCV, Bhunia_2020_CVPR} on image retrieval tasks, namely Acc.@$n$ accuracy (A@$n$), which assesses the relevancy of the top 'n' candidates retrieved.
We created 25 modification syntaxes based on templates derived from 3 lap chole videos and generated both free-drawing-based (baseline) and polygon-based (our method) reference maps. The modification syntaxes encompass three categories: 1) the relative spatial relationship of the Gallbladder, Fat, and connected tissues; 2) the size of the components; and 3) the position of the components like surgical tools placed at different places. Both the baseline and our method have the same search range. 
For each template, each method generates 9 suggestions. We asked two human raters (both have sufficient knowledge of anatomical structures in lap chole) to independently label whether the retrieved image reflects the template (blind to condition). Images that were jointly selected by both raters were considered successful suggestions. We refer to our metric as A@9 based on \cite{Bhunia_2020_CVPR} as we assess the relevancy of the top 9 candidates.

Among the 225 suggestions (9 suggestions per each of the 25 syntaxes), the baseline method achieves an A@9 of 31.11\%, while our method achieves an A@9 of 88\%. Moreover, for each syntax, an appropriate suggestion is always present using our method. In contrast, appropriate suggestions are found in only 80\% of the syntaxes when using the baseline method.

\section{Evaluation Study}
We performed an IRB-approved evaluation study with surgeons to fully evaluate all the system components and probe into surgeons' attitudes towards the educational value of the questions created using Surgment. \highlight{The study has 5 evaluation goals.}
\begin{itemize}
    \item How do surgeons compare the search-by-mask tool in identifying images from surgery videos with traditional methods? How might the retrieved images help learning?
    \item How do surgeons compare the visual feedback offered by Surgment with textual feedback or manual annotations? Is visual feedback of educational value?
    \item Are surgeons able to use Surgment to create questions and feedback of educational value?
    \item How do surgeons perceive the role of AI in facilitating the process?
    \item What challenges do surgeons experience when using Surgment and what are the design implications to develop AI-assisted methods to enhance surgical training? 
\end{itemize}

\subsection{Participants}
We recruited surgeons through mailing lists of the hospital at our university and our contacts. 11 surgeons (9 male, 2 female) from the general surgery department of 4 teaching hospitals in the United States and China participated in the study, with their demographic information displayed in Tab.\ref{tab:userstudy}. 
We originally targeted attending surgeons and senior residents (above PGY-4), who have independently performed lap chole surgeries and mentored trainees. We further extended our study invitation to three junior residents who could provide feedback from the trainees' perspectives. 
All 8 attending surgeons and senior residents have independently performed lap chole surgeries and instructed trainees in the operation room. The study sessions lasted for 60-75 minutes via Zoom. Participants were compensated with a \$50 Gift Card.
\begin{table}[h]
\begin{tabular}{ccccc}
\hline
ID  & Gender & Race      & Profession & \# Lap Chole Performed \\ \hline
P1  & Male   & Asian     & Attending        & 1000+                   \\
P2  & Male   & Cacausian & Attending        & 500+                    \\
P3  & Male   & Asian     & Attending        & 50+                     \\
P4  & Male   & Cacausian & PGY-5            & 50+                     \\
P5  & Male   & Cacausian & PGY-4            & 40+                     \\
P6  & Male   & Cacausian & PGY-4            & 40+                     \\
P7  & Male   & Cacausian & PGY-2            & 10+                     \\
P8  & Female & Cacausian & PGY-2            & 10+                     \\
P9  & Female   & Cacausian & PGY-2            & 2                      \\
P10 & Male   & Asian     & Attending        & 80+                     \\
P11 & Male   & Cacausian & PGY-5            & 50+                       \\ \hline
\end{tabular}
\caption{Demographic information of surgeon participants in the evaluation study.}
\label{tab:userstudy}
\Description[Demographic information of the evaluation study]{A table reports the Gender, Race, Profession Level, and number of Lap Chole performed for the 11 expert surgeons.}
\end{table}

\subsection{Procedure}
In the study, participants received a demo of the Surgment system, followed by three tasks: 1) Using the search-by-mask tool to find image frames, 2) Creating MCQs and providing feedback, and 3) Independently crafting 3-5 questions they deemed educationally valuable in Surgment. Participants shared screens and verbalized their thoughts during the session. 

In Task 1, participants used the search-by-mask tool for three image retrieval exercises. They were presented with a reference image and tasked to find images meeting specific criteria: ``less fat on gallbladder,'' ``zoomed-in view at gallbladder's bottom,'' ``tool centering,'' ``gallbladder stretching,'' ``undissected cystic duct and artery,'' and ``tool at gallbladder's edge'' (across two videos \cite{video1, video2}). 

We then asked for feedback on their image retrieval experience, probing if the tasks reflected their real video navigation needs, how the method compared to traditional ones, and encountered challenges.
In Task 2, participants were tasked to create three multiple-choice questions with visual feedback. Following the task, we asked participants to share when visual feedback would be beneficial for trainees and comment on its educational value in comparison to textual feedback and manual annotations. For Task 3, they were asked to freely create 3-5 questions they considered to be of high educational value. 
In the end, participants shared their views on the potential use cases of Surgment, their experience in receiving AI assistance throughout the process, the challenges they encountered, the risks they perceived, and the prospects and obstacles for using Surgment in practice.

The study recordings were transcribed and analyzed using affinity diagrams \cite{moggridge2007designing}. Two authors interpreted the transcripts and systematically grouped the interpretation notes to identify emerging themes and patterns in the data.

\section{Findings}
We present findings in response to each study goal.

\subsection{ How do surgeons compare the search-by-mask tool to identify surgery video images with traditional methods? How might the retrieved images help learning?}
\subsubsection{The search-by-mask tool helps surgeons express their requirements when searching for images} 
Most users commented that the search-by-mask tool was a mechanism that helped them express their search requirements for image frames in a surgery recording. 
They were able to describe their criterion by adjusting the masks, for example, to identify images with varying degrees of exposure for the gallbladder, different sizes of the Calot's triangle, various positions of the surgical tools, and different camera angles e.g., a zoomed-in view versus a zoomed out view.
P2 described how they retrieved an image with more exposure of the gallbladder, \textit{``The first thing that we did was just bringing the fat down where the fat is all out of the way. Yeah, I think that could be useful.''} P5 said, \textit{``I think the example of making the polygon bigger for the gallbladder and zooming in further, makes a lot of sense.''} P8 used the search-by-mask tool to locate an image where the clips had not been applied yet, by moving the polygon representing the clips out of the scene.

Surgeons commented that identifying surgery scenes through the search-my-mask tool helped them think through the educational value of the content. For example, P9 found the tool helpful for trainees to visualize what authentic surgery scenes looked like when the shape of an anatomic component was adjusted. P9 added that \textit{``I like the sketch because I'm able to visualize where it should be and also what it'll look like immediately before and after.''} P7 shared that medical students could use it to practice camera driving, as they could see how the scene would change as the camera moved: \textit{``if what you're getting at is to tell a medical student to zoom in, this is what the expected result would be as opposed to here.''
}

\subsubsection{The search-by-mask tool helps surgeons retrieve desired images efficiently}
When asked about how the search-by-mask tool compared with traditional video navigation methods, 7 out of 11 participants reported that it saved time, as it eliminated the need to watch videos and allowed for direct searching. P6 said \textit{``it's beneficial because otherwise, you would just have to keep panning through the video until you find a still frame that we wanted''.}
10 out of 11 participants considered the images retrieved using the search-by-mask tool to be of high quality and effectively meet their objectives. P9 said, \textit{``Very accurate of what I would expect and what I was envisioning when I was doing my sketch of stretching to retract the gallbladder for better visualization.''} P8 also spoke highly of the accuracy, \textit{``They seem pretty accurate from what I was looking through.''}
P1 noted that the advantage of the search-by-mask tool would be more pronounced for longer procedures, \textit{``If this method is applied to surgeries that take a longer time, like complex abdominal cancer surgeries, its advantages would be even more apparent.''}

\subsubsection{The search-by-mask tool incurs a learning curve and may add to people's cognitive load}

P3, P6, P9, and P10 prefer scrolling because they were used to this method, and found learning to use the search-by-mask tool to introduce cognitive load. P3 expressed his confusion when he used it for the first time that \textit{``at first glance, the cartoon representation of things is a little bit confusing.''} P9 said that \textit{`` I think that it would take a little bit of time and practice, ..., the more that I would use it, the easier it would become. But there's a learning curve.''} P10 mentioned, \textit{``It [the search-by-mask tool] is definitely faster, but I'd rather scroll through the video to find images because this is how I've always done it''}
P5 and P6 considered that when they were familiar with the surgery recording, they could easily recall the video content so that directly locating that image might be easier than expressing the requirement through the polygon masks. 
P6 said \textit{``I don't think of laparoscopic procedures as a bunch of pictures. If I'm looking for a still frame of a laparoscopic operation, my brain would rather think about when in this surgery that would happen and just scroll to it.''} 

\subsubsection{Surgeons describe scenarios where the search-by-mask tool is insufficient, e.g., searches that require scene rotation} 
P4 and P5 noted that in some lap chole surgeries, they needed to rotate the camera to show different angles. 
Depicting a rotation of the camera angle by modifying the polygon masks is challenging since a rotated scene can introduce substantial changes and unforeseen elements. P5 said, \textit{``I don't know how I would adjust the polygons intuitively to show me the viewpoint from the other side, which I think is obviously a pretty important part.''} He explained \textit{``You actually can't confirm that you have the critical view of safety without seeing a 3D view of the structures and moving the camera around.''}

\subsection{How do surgeons compare the visual feedback offered by Surgment with textual feedback or manual annotations? Is visual feedback of educational value?}

\subsubsection{Visual feedback enhances anatomical learning through diverse real-world cases beyond textbook diagrams.}

Most participants considered that providing visual feedback could help trainees learn operation flow and anatomy. Medical students and junior residents traditionally learn from textbooks which often portray a standardized representation of anatomical structures. However, in real surgeries, the arrangement of anatomical structures can be quite different.
P2 pointed out that \textit{``it's [visual feedbacks] additional repetitions of recognizing anatomical structures because there's the textbook diagram for what anatomy is supposed to look like, but in every single human being it looks just slightly different.''} P3 marked that \textit{``Learning from more cases is always a good thing. The more cases trainees encounter, the more proficient and familiar they become with the anatomy.''} P8 also said \textit{``We don't have a lot of question banks with images that are more operative, it's all more diagram-based. So it's definitely helpful to see what it would look like from a laparoscopic perspective.''}
Senior residents also saw value in seeing anatomy from real complex cases. 
P3 said \textit{``When I was a senior resident, I was particularly interested in complex and challenging cases, such as those where the patient has had previous open surgery and the tissues are all adhered together. The annotations would also be especially useful for me.''}

Surgery is a visual skill, and providing visual feedback aligns with surgeons' mental models. P5 shared that \textit{``a lot of medical students and residents read text to memorize the steps of the operation, ..., But actually providing the image and labeling it is obviously a lot better to show what it actually looks like when a step has been completed.''}
P7 pointed out that \textit{``surgery is a visual thing and when you put visual things into words, there's something that's always lacking. So I think the [labeled] images are helpful.''} P5 agreed that \textit{``Surgeons are a little bit more spatially and visually oriented and they are usually not abstract thinkers. So I think this appeals to our sort of visually oriented brain seeing pictures.''}

\subsubsection{Highlighting the regions is easy and makes the feedback engaging and digestible.}
Participants found highlighting critical areas in the images where trainees should focus on ensures that learners know what they should aim to achieve during procedures. P6 said, \textit{``it just is a nice way to provide clear, concise examples of the visualization that they should be looking for and should be trying to achieve.''}
Additionally, certain structures do not have clear boundaries or might be obscured by other tissues, e.g., fat, making it difficult for trainees to differentiate. 
In these cases, visual feedback highlighting an area on the scene helps.
P5 said \textit{``Seeing pictures is obviously a little bit easier to continue to study. If you have to read 10 pages of text without pictures or labeled pictures, you're probably not going to get through the whole thing.''}

\subsubsection{Surgeons need finer-grained segmentation of the images to offer highly targeted feedback.} 
Some surgeons commented that they would be able to offer higher-quality feedback if the system offered finer-grained segmentation. Some anatomical components contain subtle anatomical features, e.g., cystic duct and cystic artery. When these features are present, it requires finer-grained segmentation to focus on a small region with significant details. P7 pointed out that it is important to differentiate cystic duct and cystic artery when they are already separated: \textit{``the resolution within here you have to improve because these are actually two different structures and this thing is selecting the whole gallbladder, which at this point is dissected.''} P3 also suggested higher granularity of segmentation saying that \textit{```Connected Tissue' is too broad and encompasses critical structures like the 'cystic duct' and 'cystic artery', should be distinctly highlighted.''}

\subsection{Are surgeons able to use Surgment to create questions and feedback of educational value?}

\begin{figure*}[h]
    \centering
    \includegraphics[width=0.9\textwidth]{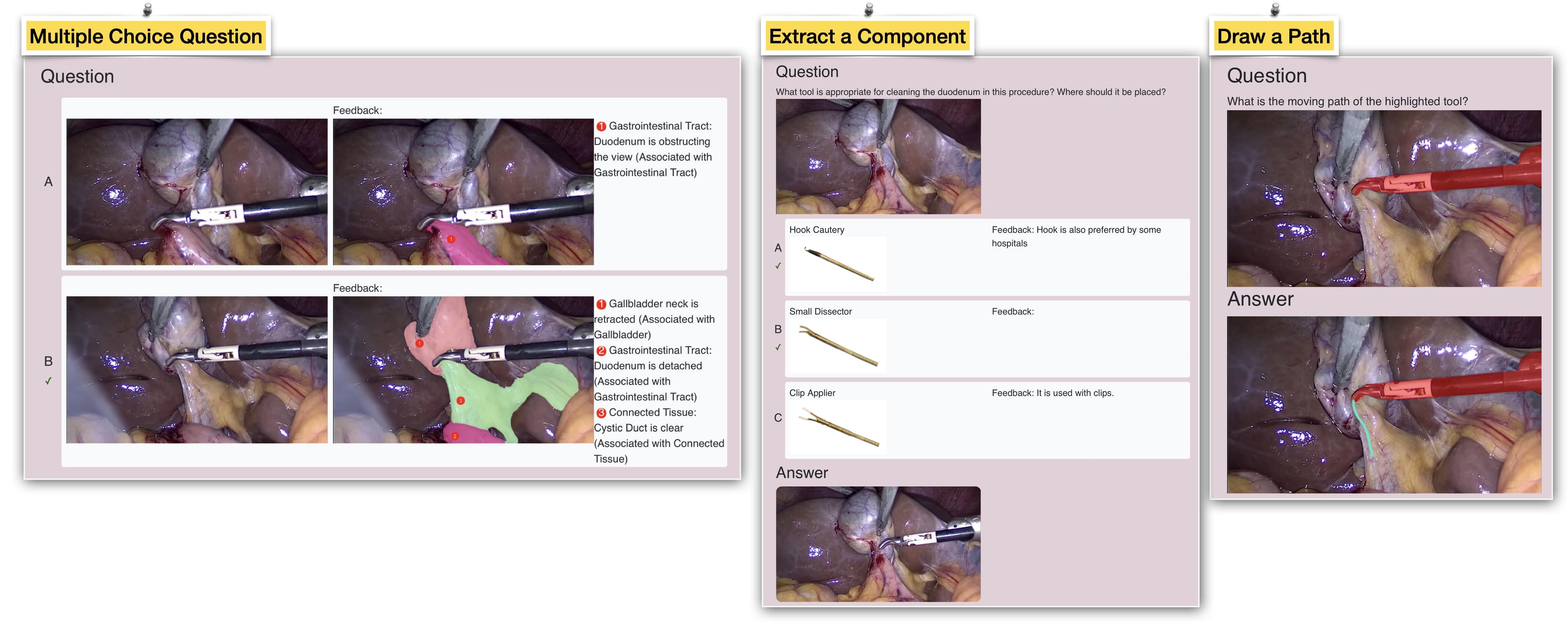}
    \caption{Example questions created by surgeons using Surgment. Q1 asks which image indicates a good timing for dissection. The surgeon also offered feedback through the highlighted regions to explain each option. Q2 asks for the right tool to detach the duodenum. Q3 asks for the movement path of the small dissector to form Calot's triangle.}
    \label{fig:example_question}
    \Description[example questions]{The left one is Q1, middle one is Q2, and right one is Q3. }

\end{figure*}

The participants considered the questions (shown in Fig.\ref{fig:example_question}) they created using Surgment to resemble the questions they had asked or were asked during an operation, citing that these questions were the most critical to ensure patient safety. P9 said \textit{``the surgical tool one [Extract a Component question type] is very useful because as you become more comfortable with the operation, the attending stops asking for the tool, and you're forced to think through which instruments should I be using here? And I think that this just sets you up to have, it forces you to start thinking about it earlier.''} P8 said \textit{``Draw a Path makes sense because there's a lot in surgery that's sort of about the movement. So it's often that we take a pause and then draw the angles or the direction of the force.''} Participants acknowledged the importance for trainees to select the correct surgical tools and apply them in the right way. Participants further mentioned that the response types enabled by Surgment would offer a new experience for surgical trainees. P2 said \textit{``Asking a trainee to draw a path is actually really interesting. I don't know that I've seen that before to be quite honest.''} 

At the same time, some surgeons raised questions on whether there existed a single solution path to these questions. For example, tool selection sometimes depends on personal preference as said by P5 that \textit{``there's no standard tool to do the next step of that operation.''} P2 agreed that \textit{``people do use all those different things [tools] for here, ..., So there are at least four different things that are theoretically useful.''} P4 also argued that in certain cases the images may not be necessary, \textit{``if it's what kind of instrument would you use here to separate the duodenum from the gallbladder, you wouldn't even need the image.''}

\subsection{How do surgeons perceive the role of AI in facilitating the process?}
\subsubsection{AI segmentation outcomes were considered to be accurate.}
All participants were notably impressed with the segmentation outcomes of the surgical scenes. P7 said, \textit{``I was pretty impressed by those color blocks getting a pretty decent screenshot of the structures we were looking at.''} and P2 agreed that \textit{``It's automatically doing that. It's incredibly impressive.''} The image inpaint function is also applauded as it removes tools and makes realistic filling. All participant was impressed by the result when they saw this function, e.g., P8 said \textit{``extract component's kind of cool.''}

\subsubsection{Surgeons perceived limitations in AI's understanding of the surgery scenes.} P5 noted that in recognizing complex structures, both surgeons and AI face difficulties. P5 considered the importance of AI to be weakened if its abilities could not go beyond humans. P5 said \textit{``For example, I already know that's the liver. I don't need the AI to tell me that it's the liver.''} P3 would want AI to help infer obscured structures (e.g., fat) in addition to parsing the current scene composition. P3 mentioned that \textit{``We sometimes cannot clearly see some complex anatomy. If AI could recognize it in these situations, that would be better. We also don't know the structure beneath the fat. If AI could predict the obscured parts, that would be something we really want.''}

\subsubsection{Surgeons raised questions about the generalizability of AI}
While acknowledging the accuracy of AI segmentation in the demonstrated examples, surgeons expressed concerns about its effectiveness in more complex procedures. They questioned whether the same level of precision would be maintained in these challenging situations. P1 expressed that \textit{``from the examples you've shown, it rarely makes mistakes. But a lap chole is a simple task. I'm not sure if it would also be applicable in much longer and more difficult surgeries, such as cardiac surgery.''} P4 said \textit{``I think it's going to be a lot more difficult for AI to figure out what's going on in those circumstances versus just the basic.''}

\subsection{What challenges do surgeons experience when using Surgment and what are the design implications to develop AI-assisted methods to enhance surgical training?}

\subsubsection{UI Challenges} 
The primary UI challenges were observed when surgeons were using the search-by-mask tool and creating visual feedback.

P4, P5, P6, and P7 reported challenges in using the search-by-mask tool due to its complexity. P4 commented \textit{``It does take a good amount of clicks to move the sketch around that. \highlight{..., because surgeons are pretty impatient,} they'll probably just scan forward.''} P5 pointed out that \textit{``clicking and dragging and resizing the polygons I think is just more arduous than searching for the actual way the operation is done.''}
Many of them shared that they preferred voice commands which felt more natural to them. P4 said \textit{``I think anything you can do to remove the extra energy required from people will be helpful.''} P5 agreed that \textit{``I mean that [voice command] might be more helpful than having to adjust manually.''}

One challenge surgeons experienced when creating feedback was that highlighting a region might obscure the details of an anatomical structure, especially when the anatomical structure was small in size. 
P8 commented on a piece of feedback she created that \textit{``Right on this one [highlighted region] that occludes the image a lot.''}. Participants suggested outlining the component instead of highlighting the entire region, or using arrows to indicate subtle structures. P4 said \textit{``you can draw arrows to show what the text is actually talking about. ..., when you're talking about the two distinct tubular structures, it's really hard to tell that that's one structure there and that's the second structure there. ''} Overall, minimal overlay on the surgery scene is preferred to maintain clarity.

\subsubsection{Surgeons suggested various ways to use Surgment in practice.}
First, many attending surgeons mentioned that they would like to use Surgment to prepare their trainees before they enter the operation room. 
P2, an attending surgeon, said \textit{`` A lot of times I'm working with a resident or a trainee and they're moving the gallbladder around and I say, ah, what we're looking for is right in there. And they clearly did not see it.''}
P10 mentioned, \textit{``Some medical students don't even know what has been done in the surgery.''} Preparing trainees with the Surgment questions could help them develop an understanding of the critical portions of the surgery, the anatomical structures, and the expectations from the attending surgeons.
Moreover, P9 emphasized the utility of offering visual feedback during an operation using a similar approach. She said, \textit{``It's really hard because more senior people can see things that I just simply can't, yet. I can't appreciate the subtleties that they can.''} Surgment can act as a tool for residents to receive precise feedback from attending surgeons.
\section{Discussion and Future Work}
The evaluation study validated surgeons’ need to enhance video-based learning. They valued Surgment's ability to semantically search for surgical scenes for teaching and favored custom visual feedback over traditional textbook diagrams. All participants successfully created high-value educational exercises and feedback, indicating Surgment's effective user interaction design. They viewed Surgment as a beneficial adjunct to current training methods, aiding trainees' pre-operative room preparation.
However, surgeons identified areas for improvement in AI-assisted educational tools. 1) They noted a learning curve with Surgment, expressing a preference for voice-command interfaces. 2) Finer segmentation was suggested for improved feedback quality. 3) While Surgment aids in understanding operational flow, anatomy, and tool handling, skills like communication, 3D perception, and decision-making require hands-on experience.
In this section, we discuss the affordances and limitations of Surgment and potential future directions for providing AI assistance to the teaching and learning of surgeries.

\subsection{The affordances and limitations of Surgment}
Surgment aims to make it easier for surgeon experts to create exercises to enhance trainees' learning. First, the formative study suggested that surgeons wanted to quickly identify surgery scenes in teaching. Surgment implements a 1) keyframe identification and a 2) search-by-mask tool to help surgeons achieve this goal. The search-by-mask tool introduces a novel interaction mechanism where surgeons can express their requirements by adjusting the shape and position of the components on a scene. The evaluation study showed that most surgeons considered this method to be more efficient than traditional video navigation techniques. Importantly, this method enabled surgeons to identify surgery scenes that met a particular teaching objective, e.g., deciding on the best timing for dissection. However, surgeons also reported a learning curve when using the tool and considered voice UI to be preferable for them.

The quiz-maker tool in Surgment enables surgeons to create three types of questions and easily offer visual feedback. In the evaluation study, surgeons appreciated the ability to highlight a region on the scene and write feedback that was attached to this visual component. Surgeons considered such annotations based on authentic surgery scenes to be more beneficial than diagrams in textbooks. Surgeons considered the two question types ``Draw a path'' and ``Extract a Component'' to be creative and helpful in teaching procedural skills and surgical tool handling skills, which are critical for lap chole surgeries. At the same time, surgeons offered suggestions to improve Surgment, e.g., providing finer-grained segmentation.  

Surgeons found Surgment beneficial for teaching surgical procedures, anatomy, and tool-handling skills. The questions created were particularly apt for junior residents who are beginning to master these skills. Some surgeons also find them useful for senior residents, especially in surgeries with complex anatomy.
Most surgeons agreed that Surgment is a valuable addition to existing training methods, helping trainees prepare for the operating room (OR). However, they noted its limitations in teaching 3D environmental understanding. Additionally, critical skills like hand-eye coordination and team collaboration, essential for surgical success, are challenging to convey through a web-based system like Surgment. 

Surgment, while aiming to make the creation of educational material easier, does not remove the need for expert surgeons to invest time and effort. Through our evaluation study, it is further confirmed that it is critical to have expert input through all stages, from identifying teachable moments to searching for surgery scenes of teaching value to composing questions using the surgery scenes and offering feedback. Many interviewed surgeons shared that the creator's expertise was essential for the questions to be useful. 

\subsection{How would surgeons like to interact and collaborate with AI?}
The core AI pillar of Surgment is the SegGPT+SAM segmentation pipeline. The segmentation outputs further support the functionalities in Surgment, including retrieving frames that surgeons want to teach and creating questions and feedback. 
The evaluation study revealed that precise scene segmentation by AI was highly valued, contributing to surgeons' positive attitudes towards Surgment. When anticipating the application of Surgment in other surgeries, the performance of the segmentation algorithm on more complex procedures is a main consideration. 
\subsubsection{Surgeons want a more intuitive and understandable interface to interact with AI}
Several surgeons considered the learning curve incurred by the search-by-mask tool to be substantial, citing that they needed to go through a series of clicks to complete a task, which was unnatural. When asked about their attitudes towards voice user interfaces, most participants agreed that they would love to verbally express their requirements. Since surgeons typically have their hands occupied while working in the OR and rely heavily on verbal communication, there's a preference for verbal interaction rather than mouse-based interactions.
Moreover, techniques could be developed to increase the learnability of the system. For example, the search-by-mask tool extracts surgery scenes that exist in the videos. If the user adjusts the polygon masks in a way that does not match any existing frames, it may not return satisfying results. The system could provide further support to scaffold the users in making modifications to the polygon masks. 
\subsubsection{Surgeons suggest ways AI should be designed to support their practice.}
Surgeons consider it helpful if AI can predict content that is not immediately visible on the scenes, such as underlying structures obscured by other anatomical components, the dynamics of internal fluids like bile, and the potential risks that may not be apparent at the surface level. Developing such features necessitates novel algorithms and human-AI interaction methods, enabling surgeons to leverage AI insights for improved operational practice and patient safety.
Our current scene segmentation effectively interprets scene composition but lacks depth perception, losing vital 3D information. It also doesn't track objects continuously across different scenes. Consequently, Surgment doesn't support identifying surgery scenes from various camera angles, a feature desired by participants. Future efforts could focus on developing techniques for object tracking across scenes. Additionally, there's a surgeon's interest in visualizing object movement, such as the motion of surgical tools or tissue deformation over time, necessitating new visualization techniques.

\subsection{Generalizability of the segmentation pipeline}
We consider the SegGPT+SAM segmentation pipeline generalizable to other laparoscopic procedures. It requires users to annotate a modest number of images, approximately 15-20 for each anatomical structure in a 10-minute video, a task easy and feasible for medical professionals with relevant knowledge. This pipeline's versatility allows for its application in segmenting scenes from laparoscopic surgeries even in the absence of public datasets, which are scarce due to patient privacy concerns. In user studies, we sought surgeons' views on using Surgment for different surgical types. Participants unanimously agreed that the question formats and visual feedback features of Surgment are highly suitable for other laparoscopic or robotic procedures, such as inguinal and umbilical hernias.

\subsection{Implications on supporting surgery teaching and learning}
The evaluation study revealed surgeons' ideas for utilizing Surgment in various ways. A promising avenue is developing AR applications using the segmentation pipeline for real-time teaching in the OR. Surgery's visual nature can constrain OR communication, especially in explaining spatial relationships. Existing approaches have not demonstrated success in addressing intra-operative communication challenges. For example, trainees find telepointers do not convey sufficient information \cite{semsar2019telepointer}, and that constructing 3D anatomical models can be expensive and lose real-world fidelity \cite{datta2012simulation}. Extending Surgment to an AR environment might facilitate feedback offering and learning in the operation room. 
Focusing on user interactions for creating questions and feedback, surgeons also saw potential in the search-by-mask tool for video navigation and learning. Inspired by Surch \cite{kim2023surch}, we envision extending Surgment to allow users to retrieve and compare scenes from multiple surgery videos, further enriching the learning experience.

\subsection{Image-based question creation}

Prior works on question generation are primarily text-based \cite{Wang2019, Yeckehzaare2020, Das2021, Kurdi2020, Wang2022, Wang2018, Majumder2014, Majumder2015, Shen2018OnTG, Yuan2019AutomaticGO, Zhu2021}. Automatic visual question generation techniques for educational purposes often produce simple questions, e.g., counting objects in a scene \cite{Xie2021, fan2018}. Techniques to generate visual questions that target higher-order thinking skills are limited. Recent work on question generation \cite{lu2023readingquizmaker, wang2022towards} also emphasized the importance of incorporating expert input to ensure the quality of the produced questions. Surgment offers another example of the importance of involving human experts at all stages in the question-creation process. The image-based question and response types implemented in Surgment are generalizable to other contexts for people to learn visual procedures and the placement of objects, and to contrast visual elements on scenes.

\section{Conclusions}
We introduce Surgment, a system to help surgeon experts efficiently create image-based questions and feedback in support of surgical trainees' learning of operation procedures, anatomy, and instrument handling skills.
Surgment is powered by a surgery scene segmentation pipeline, which combines two state-of-the-art few-shot learning models SAM and SegGPT. 
Surgment streamlines the process of pinpointing frames of interest by enabling users to express their requirements by adjusting the size, shape, and positions of the components on a scene. Moreover, Surgment supports users to create diverse types of questions and visual feedback targeting operation flow, anatomy, and surgical tool handling skills.
In the evaluation study, all surgeons were able to successfully create exercises and feedback that they considered to be of high educational value, suggesting that the user interactions in Surgment were effectively designed. Surgeons considered Surgment to be a nice complement to existing training approaches and could help trainees prepare before they enter the operation room. The study identified areas for future work, including exploring voice user interfaces, improving the granularity of the segmentation outputs, providing better visualization techniques to show the movements of objects over time, and predicting invisible components on the scenes.

\bibliographystyle{ACM-Reference-Format}
\bibliography{sample-base}

\end{document}